\documentclass[submission, Phys]{SciPost}

\usepackage{graphicx}
\usepackage{bm}
\usepackage{color}
\usepackage{physics}
\usepackage{amsfonts}
\usepackage{amssymb}
\usepackage{amsthm}
\usepackage{amsmath}
\usepackage{mathrsfs}
\usepackage{subfig}

\begin{document}

\begin{center}{\Large \textbf{
  Antagonistic interactions can stabilise fixed points in heterogeneous linear dynamical systems
}}\end{center}

\begin{center}
Samuel Cur\'{e}\textsuperscript{1},
Izaak Neri\textsuperscript{2},
\end{center}

\begin{center}
{\bf 1} Okinawa Institute of Science and Technology, 1919-1 Tancha, Onna-son, Okinawa 904-0495, Japan
\\
{\bf 2} Department of Mathematics, King’s College London, Strand, London, WC2R 2LS, UK
\\
*samuel.cure@oist.jp
\end{center}

\begin{center}
\today
\end{center}


\section*{Abstract}
{\bf
We analyse the stability of large, linear dynamical systems of variables that  interact through a fully connected random matrix and have inhomogeneous growth rates.  We show that in the absence of correlations between the coupling strengths, a system with interactions  is always less stable than a system without interactions. Contrarily to the uncorrelated case, interactions that are  antagonistic, i.e., characterised by  negative correlations, can stabilise linear dynamical systems.  In particular,  when the strength of the interactions is not too strong,  systems with antagonistic interactions are more stable than systems without interactions. 
These results are obtained with an exact theory for the spectral properties of fully connected random matrices with diagonal disorder.
}

\vspace{10pt}
\noindent\rule{\textwidth}{1pt}
\tableofcontents\thispagestyle{fancy}
\noindent\rule{\textwidth}{1pt}
\vspace{10pt}

\section{Introduction}
\label{sec:intro}
We consider a dynamical system described by $n$ variables $x_j\in\mathbb{R}$ that are labeled by  indices $j=\left\{1,2,\ldots,n\right\} = [n]$ and where  $t\in \mathbb{R}^+$ is the time index.  The evolution in time of the variables  $x_j(t)$ is described by a set of randomly coupled, linear differential equations of the form 
\begin{equation}
\partial_t x_j(t) = \sum^{n}_{k=1}A_{jk} \: x_k(t) , \label{eq:1}
\end{equation}
where the $A_{jk}$ are the entries of a  random matrix $\mathbf{A}$ of dimension $n\times n$.    
The  fixed point $\vec{x} = 0$  of the set of Eqs.~(\ref{eq:1}) is stable when all eigenvalues of $\mathbf{A}$ have negative real parts.  On the other hand, if there exists at least one eigenvalue with a positive real part, then the fixed point is unstable.   

Differential equations of the form Eq.~(\ref{eq:1}) appear in  linear stability analyses  of  complex systems described by    nonlinear differential equations of the form $\partial_t \vec{y}(t) = \vec{f}(\vec{y}(t))$ where $\vec{y} = (y_1,y_2,\ldots,y_n)$. For example, in theoretical ecology ecosystems are modelled with Lotka-Volterra equations, where the variable $\vec{y}$ quantify the population abundances of the different species in the population~\cite{may2019stability}.   Other examples are models for neural networks, for which $\vec{y}$ represents the neuronal firing rates or the membrane potentials \cite{chaos,kadmon2015transition,ahmadian2015properties}, and models of economies \cite{quirk1965qualitative}, for which $\vec{y}$ represents economic variables such as the prices of goods.    If the differential system determined by $\vec{f}$ admits a fixed point, defined as $\vec{f}(\vec{y}^\ast)=0$, then the dynamics of  $\vec{x} = \vec{y}-\vec{y}^\ast$  near the fixed point is given by Eq.~(\ref{eq:1}), where $\mathbf{A}$ is the Jacobian of  $\vec{f}$. The linear stability  of a complex system that settles in a fixed point   state is thus determined by the real part of the {\it leading eigenvalue} $\lambda_1$, which is defined as an eigenvalue of the Jacobian matrix $\mathbf{A}$ that has   the largest real part. 

 To study complex systems, Wigner \cite{wigner1958distribution}, Dyson \cite{dyson1962statistical} and May \cite{may1972will}, among others, suggested to study    random matrices  $\mathbf{A}$,  and  the task at hand is then to determine the real part of the leading eigenvalue as a function of  the parameters that define the random matrix ensemble.     Although one should be careful in drawing conclusions   about the dynamics of nonlinear systems from the study of  randomly coupled linear differential equations,  random matrix theory has the advantage of providing   analytical insights about the influence of interactions on linear stability.     In fact,  linear stability analyses with random matrix theory have been used to    study       the onset of chaos in random neural networks \cite{chaos, kadmon2015transition,ahmadian2015properties},   the stability of ecosystems modelled by Lotka-Volterra equations with random interactions \cite{may1972will, allesina2012stability, mougi2012diversity, mougi2014stability, biroli2018marginally, roy2019numerical, 10.21468/SciPostPhys.12.1.013}, economies \cite{moran2019may}, or   gene regulatory networks \cite{guo2021exploring}.  Moreover, although traditionally random matrix models are fully connected,  recently exact results have been derived for the  stability of linear models defined   on complex networks  \cite{neri2020linear, tarnowski2020universal, mambuca2020dynamical}.  
 

So far,  stability analyses for randomly coupled, linear dynamical systems have mainly focused on matrices $\mathbf{A}$ with  diagonal entries, which we also call the growth rates,  that are fixed to a constant value $d$, i.e., \cite{may1972will}
\begin{equation}
A_{jk} = \frac{J_{jk}}{\sqrt{n}}(1-\delta_{j,k}) + d \delta_{j,k},  \label{eq:dconstant}
\end{equation}
where $\delta_{j,k}$ is the Kronecker delta function, and where the coupling strengths $J_{jk}$ are random variables drawn from a certain distribution.   Following Refs.~\cite{allesina2012stability, cicuta2016random}, we   consider the case where the pairs of random variables $(J_{jk},J_{kj})$ are  independent and identically (i.i.d.) distributed random variables drawn from a distribution with
\begin{equation}
\langle J_{ij}\rangle = 0, \quad \langle J^2_{ij} \rangle = \sigma^2, \quad {\rm and} \quad \left\langle J_{i j} J_{j i}\right\rangle = \tau \sigma^2 ,  \label{momentsOfMatrixJ}
\end{equation} 
where the variance $\sigma^2$ of the entries $J_{ij}$  quantifies the strength of the interactions, and   $\tau\in[-1,1]$ is the Pearson correlation coefficient  between the variables $J_{jk}$ and $J_{kj}$.       The sign of the parameter $\tau$ is important in theoretical ecology as it determines the nature of the trophic interactions between two species.   If the interactions are on average competitive ($J_{ij}<0$ and $J_{ji}<0$) or mutualistic ($J_{ij}>0$ and $J_{ji}>0$), then $\tau>0$.   On the other hand, if the interactions are on average antagonistic ($J_{ij}>0$ and $J_{ji}<0$ or $J_{ij}<0$ and $J_{ji}>0$),  then $\tau<0$ \cite{allesina2012stability, mougi2012diversity, mougi2014stability, mambuca2020dynamical}.   In theoretical ecology, antagonistic interactions are also called predator-prey interactions as  they describe trophic interactions between two species for which one predates on the other.

The  leading eigenvalue of random matrices of the form (\ref{eq:dconstant}) is given by \cite{may1972will,girko1986elliptic, sommers1988spectrum, PhysRevLett.60.1895}
\begin{equation}
    \text{Re}(\lambda_1)= \sigma(1+\tau)+d. \label{eq:leading1}
\end{equation}   
It follows from  Eq.~(\ref{eq:leading1})  that in the case of homogeneous relaxation rates  $d<0$ is required for a linear system to be stable.   Hence, when the diagonal entries of $\mathbf{A}$ are fixed to a constant value $d$, then  interactions  $J_{jk}$  always destabilise fixed points in large dynamical system.

In the model given by Eq.~(\ref{eq:dconstant}) it holds that in the absence of interactions ($J_{ij}=0$) either all variables are stable  (when $d<0$) or all variables are unstable (when $d>0$).  In this paper, we relax this condition and consider random matrix models with growth rates $A_{jj} = D_j$ that fluctuate from one variable to the other.   In the symmetric case ($\tau=1$), such random matrices are called deformed Wigner matrices \cite{Pastur, lee2016extremal, mergny2021stability}  and  in this case a functional equation that determines the spectral distribution  in the limit of large $n$ has been derived by Pastur in  Ref.~\cite{Pastur}.   Another case  that has been studied  in the literature is when $\mathbf{A}$ is the  adjacency matrix of a random directed graph with diagonal disorder~\cite{neri2016eigenvalue, neri2020linear, tarnowski2020universal}, which corresponds  in the dense limit with $\tau=0$ \cite{khoruzhenko1996large}, and for which a simple equation for the  boundary of the spectrum as a function of the distribution of diagonal matrix entries   has been derived. 

On the other hand, in the present paper we focus on the case of heterogeneous relaxation rates $D_j$ with negative $\tau$, which is, as discussed, of particular interest for ecology.   This case has been studied signficantly less in the literature, a notable exception being Ref.~\cite{barabas2017self}.   Here, apart from completing the theory of Ref.~\cite{barabas2017self} by  deriving analytical results for eigenvalue outliers, which are important when considering the leading eigenvalue, we also show that  for negative $\tau$ the leading eigenvalue can be negative, even if  a finite fraction of the relaxation rates  $D_j$ are positive. The latter finding, not discussed in Ref.~\cite{barabas2017self}, implies   that antagonistic interactions can stabilise linear  systems when the interactions are neither too weak nor too strong,  even if a finite fraction of the variables are unstable in isolation, and this constitutes the main result of this paper.

The paper is organised as follows.  In Sec.~\ref{sec:2} we define the model that we study, which is a fully connected random matrix with  diagonal disorder.   In Sec.~\ref{sec:3}, we discuss the cavity method, which is a method from theoretical physics that we use to study the model in the limit of infinitely large random matrices.  In Sec.~\ref{sec:3x}, we present the main results for the boundary  for the spectrum of fully connected matrices with diagonal disorder and in Sec.~\ref{sec:outlier} we present analytical results for the eigenvalue outlier.  In Sec.~\ref{sec:4} we use the obtained theoretical results to derive phase diagrams for the linear stability of fixed points.    We end the paper with a discussion in Sec.~\ref{sec:5}.   The paper also contains  a few appendices where we present details about the mathematical derivations.

\section{Fully connected random matrices with diagonal disorder}\label{sec:2}
We consider the random matrix  model 
\begin{equation}
A_{jk} = \frac{J_{jk}}{\sqrt{n}}(1-\delta_{j,k}) + \frac{\mu}{n} + D_j\delta_{j,k},  \label{eq:AD}
\end{equation}
where  the (off-diagonal) pairs $(J_{ij},J_{ji})$ are i.i.d.~random variables  drawn from a joint distribution $p_{J_1,J_2}$ with moments   as specified in the Eqs.~(\ref{momentsOfMatrixJ}),  where the diagonal elements $D_j$   are i.i.d.~random variables drawn from a distribution $p_D$, and where $\mu\in \mathbb{R}$ is a constant shift of the matrix elements.      
Note that without loss of generality we have set $\langle J_{ij}\rangle = 0$, as a nonzero average value can be incorporated into the parameter $\mu$.

As will become clear later,  just as is the case for the circular law  \cite{bai2008methodologies, tao2012topics}, in the limit of $n\gg 1$ the boundary of the spectrum of $\mathbf{A}$ is a deterministic  curve  in the complex plane that depends on the distribution  $p_{J_1,J_2}$ of $(J_{ij},J_{ji})$  only through its first two moments given in Eq.~(\ref{momentsOfMatrixJ}), and hence we will not need to specify $p_{J_1,J_2}$.   On the other hand, the boundary of the spectrum of $\mathbf{A}$ depends in a nontrivial way on the distribution  $p_D$, and therefore it will be interesting to  study the effect that the shape of    $p_D$ has on the leading eigenvalue.    Due to the constant shift $\mu$, the spectrum may also contain a single (deterministic) eigenvalue outlier in the limit of large $n\gg 1$~\cite{tao2013outliers, o2014low, neri2016eigenvalue, neri2020linear}.

In the  special case when $p_D(x) = \delta(x-d)$ we recover the model given by  Eq.~(\ref{eq:leading1}).      A  more interesting case is when the growth rates $D_j$ are heterogeneous, and arguably the most simple model for heterogeneous growth rates considers that the $D_j$ can take two possible values, yielding  a   bimodal distribution 
\begin{equation}
    p_D(x)= p\: \delta(x-d_-)+(1-p)\:\delta(x-d_+),\label{eq:bimodalx} \end{equation} 
    with $d_-<0$,  $d_+>0$, and $p\in [0,1]$, and where $\delta(x-d)$ denotes the Dirac delta distribution.     In this example,    a fraction $(1-p)$ of variables $x_j$ are unstable in the absence of interactions    ($\sigma^2=0$).  
 We  also consider cases where $p_D$ is a continuous distribution.   One example of a continuous distribution is the  uniform distribution defined on an  interval $[d_-,d_+]$, i.e.,
 \begin{equation}
    p_D(x)= \left\{ \begin{array}{ccc} 0 &{\rm if}& x \notin [d_-,d_+] ,\\ \frac{1}{d_+-d_-} &{\rm if}& x\in [d_-,d_+]. \end{array}\right.\label{eq:uniform}\end{equation} 
Since the uniform distribution is supported on a bounded set, we will also consider an example  for which $p_D$ has unbounded support, namely, we will consider the   Gaussian distribution 
 \begin{equation}
     p_D(x) = \frac{1}{\sqrt{2\pi }} e^{-\frac{x^2}{2}}\label{eq:gauss}
 \end{equation}
 with zero mean and unit variance.  
 
    The main  question we  address in this paper  is whether the interaction variables $J_{ij}$ can stabilise a linear dynamical system even when a finite fraction of variables are unstable in the absence of interactions, i.e., a finite fraction of species $i\in[n]$ have a positive growth rate $D_i$.  In other words, we ask whether it is possible to have  $\Re \lambda_1<0$ even when there exists a value $d>0$ such that $p_D(d)>0$.

\section{Cavity method for the empirical spectral distribution of infinitely large  matrices}\label{sec:3}
We determine the leading eigenvalue $\lambda_1$ in the   case of $\mu=0$, when the spectrum of $\mathbf{A}$ has no outliers in the limit of $n\rightarrow \infty$.   

In this case, the  leading eigenvalue of the adjacency matrices $\mathbf{A}$ is determined by  the empirical spectral distribution $\rho$ of the eigenvalues $\lambda_j$ of $\mathbf{A}$, defined by 
\begin{equation}
    \rho(z)=\lim_{n\rightarrow \infty}\frac{1}{n}\left\langle\sum_{j=1}^{n} \delta\left(x-\operatorname{Re}( \lambda_{j})\right) \delta\left(y-\operatorname{Im} (\lambda_{j})\right)\right\rangle
    \label{spectralDensityDefinition}
\end{equation}
for all $z = x + {\rm i}y\in\mathbb{C}$.     The spectral distribution determines the leading eigenvalue  of the continuous part of the spectrum through
\begin{equation}
\lambda_1 =  {\rm argmax}_{\left\{z\in \mathbb{C}:\rho(z)>0\right\}} \Re (z).   \label{eq:leadingEigenvalueRho}
\end{equation}
Equation (\ref{eq:leadingEigenvalueRho}) holds as long as  the spectrum of $\mathbf{A}$ does not have eigenvalue outliers \cite{neri2016eigenvalue, neri2020linear}, which  for the model defined in Sec.~\ref{sec:2} is the case as long as $\mu=0$ ~\cite{neri2016eigenvalue, neri2020linear}.  

The convergence in Eq.~(\ref{spectralDensityDefinition}) should be understood as weak convergence \cite{tao2012topics}, which implies that the average of any bounded and continuous function $f(z)$ defined on the complex plane converges in the limit of large $n$ to $\int_{\mathbb{C}}{\rm d}z\: \rho(z) f(z)$.  Also, we can drop the average in the right-hand side of Eq.~(\ref{spectralDensityDefinition}) as  the spectral distribution converges almost surely and weakly to $\rho$ \cite{tao2012topics}, and hence also the leading eigenvalue $\lambda_1$ as defined in Eq.~(\ref{eq:leadingEigenvalueRho}) is a deterministic variable for large values of $n$.  

The limiting distribution $\rho$ of random matrix models as defined in  Sec.~\ref{sec:2} have been studied before in several special cases.  Notably, for the symmetric case with $\tau=1$ Pastur derived a functional equation that determines $\rho$ \cite{Pastur}.  Recently,  the symmetric case was  revisited in \cite{mergny2021stability}, and  in that reference also the large deviations of $\lambda_1$ were computed in the  case when the matrix entries $J_{ij}$ are drawn from a Gaussian distribution; note that large deviations are not universal and depend on the statistics of $(J_{ij},J_{ji})$ as determined by the distribution $p_{J_1,J_2}$.      In the  case when $\tau=0$ and $p_D$ is a bimodal distribution  the spectral distribution $\rho$ has been determined in Refs.~\cite{feinberg1997non, hikami1998density} and the  $\tau=0$ case for general $p_D$ has been considered in \cite{khoruzhenko1996large}.   For random directed graphs with a prescribed distribution of indegrees and outdegrees, which corresponds with the case $\tau=0$ in the limit of large mean degrees, a simple equation was derived for the boundary of the spectrum in Refs.~\cite{neri2016eigenvalue, neri2020linear, tarnowski2020universal}.     Lastly, Ref.~\cite{barabas2017self} obtained analytical results for the spectrum when $\tau<0$ and $\mu=0$.

We determine the spectral density $\rho(z)$ from the resolvent of the matrix $\mathbf{A}$, which can be determined with the cavity method \cite{rogers2009cavity, metz2019spectral}.  The resolvent   is defined as
\begin{eqnarray}
\mathbf{G}(z) = \left(z\mathbf{1}_n-\mathbf{A}\right)^{-1} ,\quad z\notin \left\{\lambda_1,\lambda_2,\ldots,\lambda_n\right\}, \label{eq:11}
\end{eqnarray}
where    $\mathbf{1}_n$ is the identity matrix of size $n$.  
The spectral distribution can be expressed  in terms of the diagonal elements of the resolvent by \cite{feinberg1997non}
\begin{eqnarray}
\rho(z) = \lim_{n\rightarrow \infty} \frac{1}{\pi n} \partial^\ast {\mathrm{Tr}}\mathbf{G}(z), \quad {\rm where} \quad \partial^\ast = \frac{1}{2}\frac{\partial}{\partial x} +  \frac{{\mathrm i}}{2}\frac{\partial}{\partial y}. \label{eq:12}
\end{eqnarray}

For non-Hermitian matrices, the eigenvalues are in general complex-valued, and therefore in the limit of $n\rightarrow \infty$ we cannot get $\rho(z)$ from ${\mathrm{Tr}}\mathbf{G}(z)$ \cite{PhysRevLett.60.1895}.  To overcome this, we use the  Hermitization method \cite{feinberg1997non}  that considers the enlarged $2n\times 2n$ matrix  
\begin{eqnarray}
\mathbf{H} = \left(\begin{array}{cc} \eta \mathbf{1}_n & z\mathbf{1}_n-\mathbf{A} \\ z^\ast \mathbf{1}_n - \mathbf{A}^T & \eta \mathbf{1}_n \end{array}\right), \label{eq:H}
\end{eqnarray}
where we have introduced a regulator $\eta$ that keeps all quantities well-defined in the limit of large $n$,  
where $\mathbf{A}^T$ is the transpose of the matrix $\mathbf{A}$, and  where $z^\ast$ is the complex conjugate of $z$.  The inverse of the matrix $\mathbf{H}$ is 
\begin{eqnarray}
\mathbf{H}^{-1} = \left(\begin{array}{cc} \frac{\eta}{\eta^2 \mathbf{1}_n -       \mathbf{I}_{\rm l} } & -(\eta^2 \mathbf{1}_n - \mathbf{I}_{\rm l})^{-1} \left(z\mathbf{1}_n-\mathbf{A}\right) \\   -\left(z^\ast \mathbf{1}_n - \mathbf{A}^T\right) \left(\eta^2 \mathbf{1}_n - \mathbf{I}_{\rm l}\right)^{-1} &  \frac{\eta}{\eta^2\mathbf{1}_n - \mathbf{I}_{\rm r}} \end{array}\right), 
\end{eqnarray}
where 
\begin{equation}
    \mathbf{I}_{\rm l} = \left(z\mathbf{1}_n-\mathbf{A}\right) \left(z^\ast \mathbf{1}_n - \mathbf{A}^T\right) \quad  {\rm and}  \quad   \mathbf{I}_{\rm r} =   \left(z^\ast \mathbf{1}_n - \mathbf{A}^T\right)\left(z\mathbf{1}_n-\mathbf{A}\right).
\end{equation}
In the limit $\eta\rightarrow 0$, we obtain 
\begin{equation}
     \mathbf{H}^{-1} = \left(\begin{array}{cc} \mathbf{0}_n & \mathbf{G}^{\rm T}(z^\ast) \\ \mathbf{G}(z)&  \mathbf{0}_n  \end{array}\right) - \eta  \left(\begin{array}{cc} \mathbf{I}^{-1}_{\rm l} &  \mathbf{0}_n\\  \mathbf{0}_n&  \mathbf{I}^{-1}_{\rm r} \end{array}\right)  + \eta^2 \left(\begin{array}{cc} \mathbf{0}_n&   \mathbf{I}^{-1}_{\rm l}\mathbf{G}^{\rm T}(z^\ast)\\  \mathbf{G}(z)\mathbf{I}^{-1}_{\rm l}&  \mathbf{0}_n \end{array}\right)  +  O(\eta^3), \label{eq:HInverse}
\end{equation}
where $\mathbf{0}_n$ is the matrix with zero entries. 
Hence, combining Eqs.~(\ref{eq:12}) and (\ref{eq:HInverse}), we find that 
\begin{equation}
\rho(z) = \lim_{n\rightarrow \infty}  \lim_{\eta\rightarrow 0} \frac{1}{\pi n}  \partial^\ast {\rm Tr}_{21}\mathbf{H}^{-1} ,  \label{eq:rhoTr21}
\end{equation}
where ${\rm Tr}_{21}$ is a block trace over the diagonal of the lower-left block of $\mathbf{H}^{-1}$. 

Defining  the $jk$-th block of the generalized resolvent as
\begin{equation}
\mathsf{G}_{j k}=\left(\begin{array}{cc}
{\left[\mathbf{H}^{-1}\right]_{j, k}} & {\left[\mathbf{H}^{-1}\right]_{j, k+n}} \\
{\left[\mathbf{H}^{-1}\right]_{j+n, k}} & {\left[\mathbf{H}^{-1}\right]_{j+n, k+n}}
\end{array}\right), \label{eq:14}
\end{equation}
  the spectral distribution (\ref{eq:rhoTr21})  can be written as  \cite{metz2019spectral} 
\begin{equation}
\rho(x, y)=\lim _{n \rightarrow \infty}\lim _{\eta \rightarrow 0}  \frac{1}{\pi } \partial^\ast g_{21}, \label{eq:rhoZ}
\end{equation}
where  
\begin{equation}
\mathsf{g} = \left(\begin{array}{cc}
 g_{11}& g_{12} \\ 
 g_{21} & g_{22}
\end{array}\right) =  \frac{1}{n}\sum^n_{j=1} \mathsf{G}_{jj} . \label{eq:16}
\end{equation}
 In Appendix~\ref{AppA}, we use the cavity method to derive a selfconsistent equation for the matrix $\mathsf{g}$ at fixed $\eta$ in the limit of $n\gg 1$, viz., 
\begin{equation}
 \left(\begin{array}{cc}
 g_{11}& g_{12} \\ 
 g_{21} & g_{22}
\end{array}\right)=\left \langle \left(\begin{array}{cc}
\eta -\sigma^2 g_{22} & z-D-\tau\sigma^2 g_{21} \\ 
z^{*}-D-\tau \sigma^2g_{12} & \eta - \sigma^2g_{11}
\end{array}\right)^{-1}\right \rangle_D ,
\label{integralResolventEquation}
\end{equation}
where $\langle \dots \rangle_D$ denotes  the average over the distribution $p_D$.   

Note that to derive (\ref{integralResolventEquation})  we have determined  $\mathsf{g}$ at finite values of $\eta$  in the limit of large $n$, and afterwards we take the limit of $\eta\rightarrow 0$.   Hence,  we interchange the two limits in Eq.~(\ref{eq:rhoZ}), which is not evident as the leading order, correction terms in Eq.~(\ref{eq:HInverse}) at large values of $n$ and small values of $\eta$ intertwine the two limits.    Demonstrating that these two limits can be interchanged constitutes   the main challenge in rigorous approaches to non-Hermitian random matrix theory, see e.g.~Refs.~\cite{bai2010spectral, tao2012topics, bordenave2012around, wood2012universality, cook2018non, cook2019circular}.   This involves   bounding the rate at which the   least singular value of $z\mathbf{1}_n-\mathbf{A}$ converges to zero for large values of $n$, as the correction terms in (\ref{eq:HInverse}) depend on the inverse of the matrices $\mathbf{I}_{\rm l}$ and $\mathbf{I}_{\rm r}$.  In this paper, we use the theoretical physics approach, i.e., we  exchange the two limits in good faith and then corroborate theoretical results with direct diagonalisation results.  In the next section, we use the Eq.~(\ref{integralResolventEquation}) together with Eq.~(\ref{eq:rhoZ}) to determine the boundary of the support set of $\rho$ in the complex plane.  

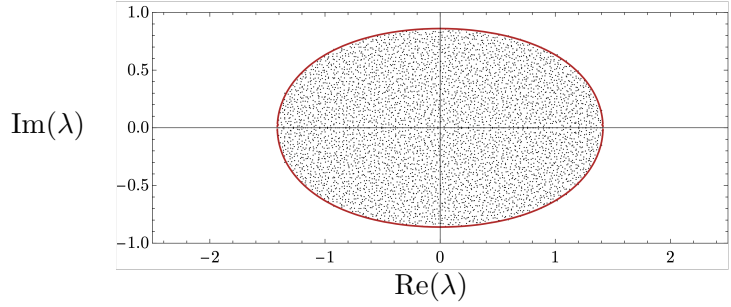
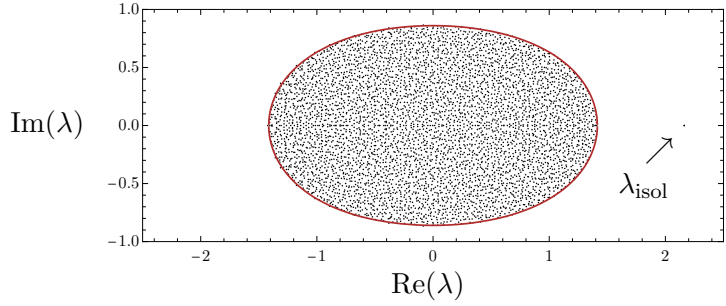
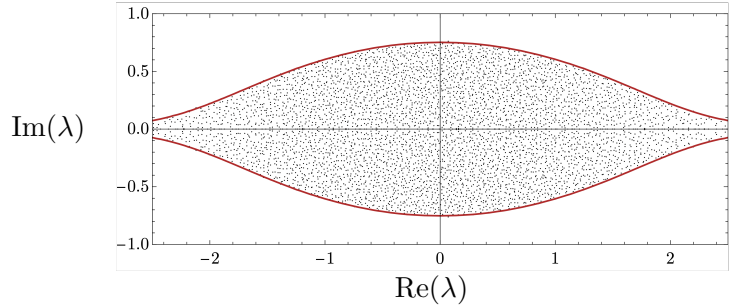
\begin{figure*}[h!]
     \centering
     \subfloat[Uniform distribution $p_D$ as defined in Eq.~(\ref{eq:uniform}) with $d_+=1$ and $d_-=-1$; the parameter $\mu=0$.]{ \includegraphics[scale=0.072]{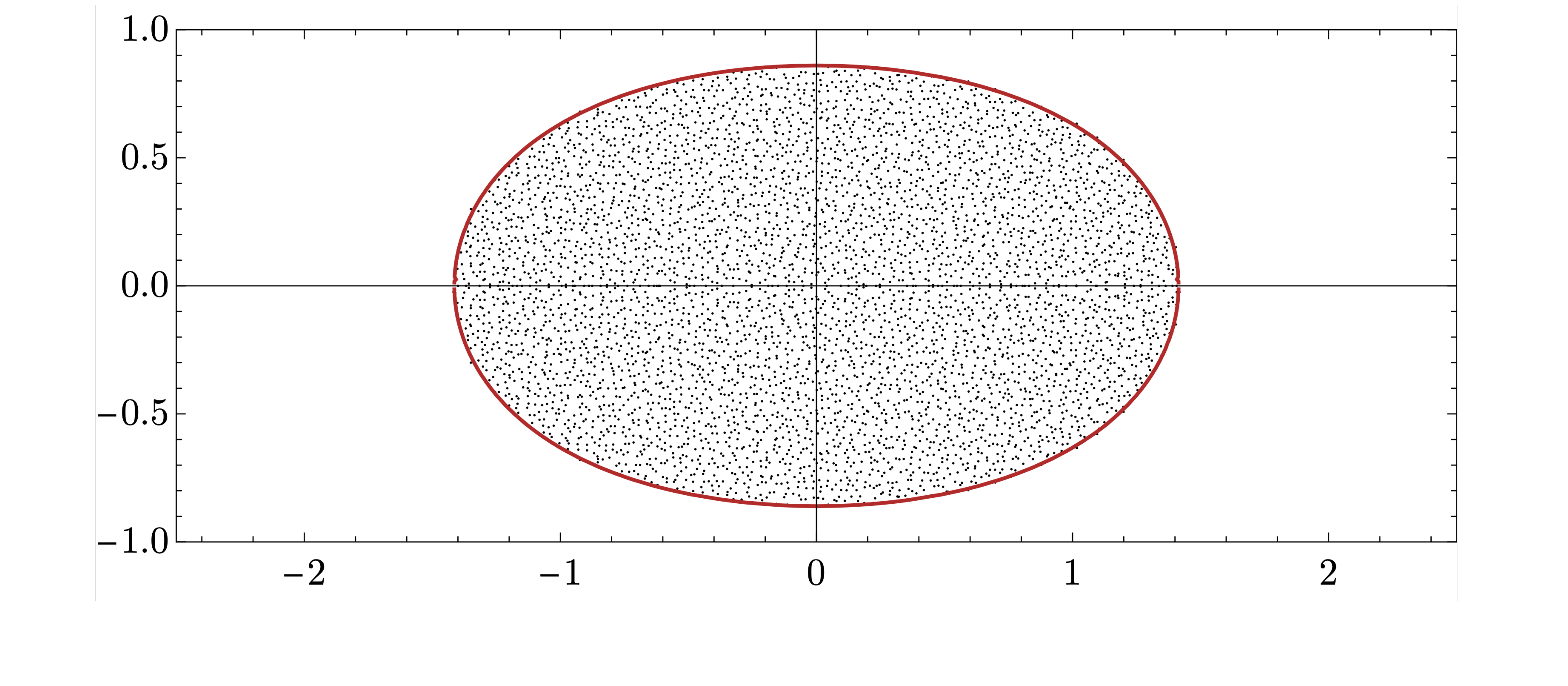}}
     \put(-290,66){$\Im(\lambda)$}
     \put(-145,5){$\Re(\lambda)$}
     \vspace{0.2mm}
    \subfloat[Uniform distribution $p_D$ as defined in Eq.~(\ref{eq:uniform}) with $d_+=1$ and $d_-=-1$; the parameter $\mu=2$. The arrow points at the eigenvalue outlier]{ 
    \includegraphics[scale=0.56]{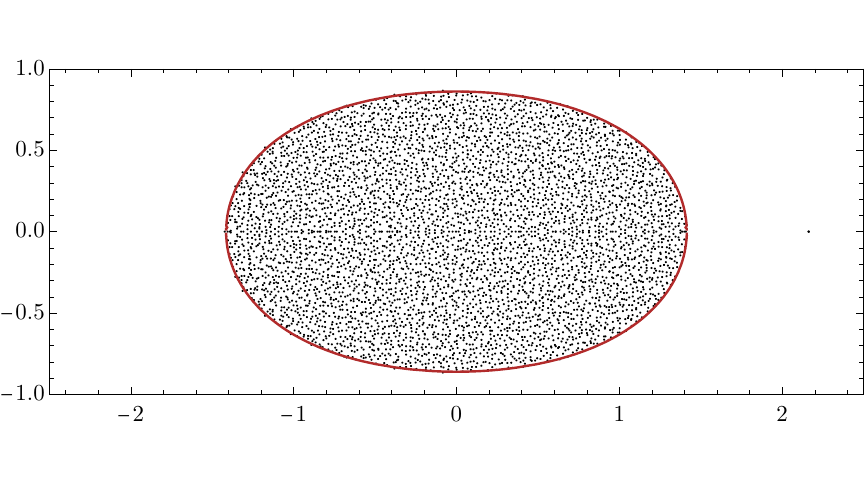}%
     \put(-40,43){$\lambda_{\rm isol}$}
    \put(-30,56){$\nearrow$}
    }
     \put(-270,66){$\Im(\lambda)$}
     \put(-126,5){$\Re(\lambda)$}
     \vspace{0.2mm}
     \subfloat[Gaussian distribution $p_D$ as defined in Eq.~(\ref{eq:gauss}); the parameter $\mu=0$.]{\includegraphics[scale=0.072]{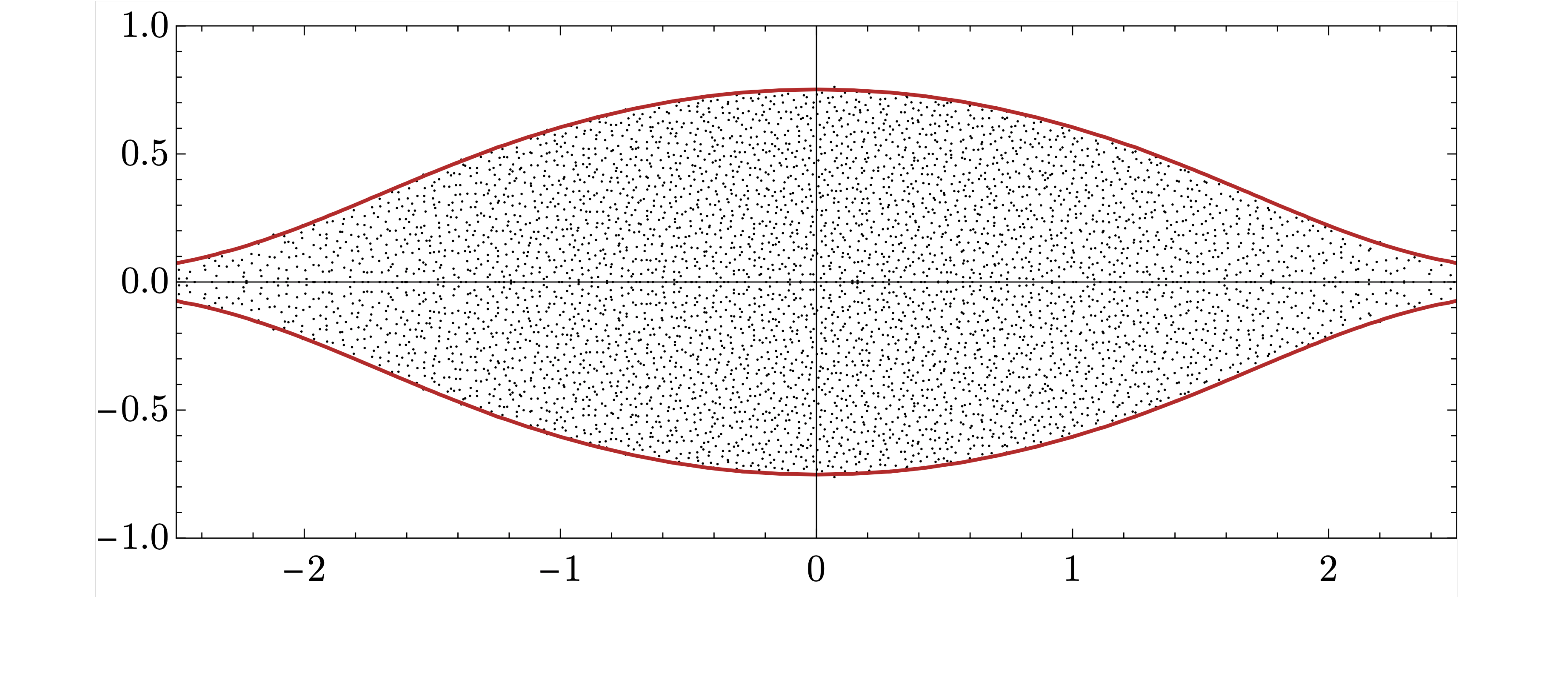}}
     \put(-290,66){$\Im(\lambda)$}
     \put(-145,5){$\Re(\lambda)$}
     \caption{Spectra of three random matrices $\mathbf{A}$ as defined in Eq.~(\ref{eq:AD}) for the uncorrelated case $\tau=0$ and with diagonal elements that are independently drawn from  a  uniform distribution [Panel (a) and Panel(b)] or  a Gaussian distribution [Panel (c)].   Markers denote the eigenvalues of a random matrix of size $n=3000$ and with off-diagonal elements $A_{ij} = J_{ij} + \mu/n$, where the $J_{ij}$  are  drawn independently from a Gaussian distribution with zero mean and unit variance and where $\mu$ is as given in the subfigure captions.   The red solid line denotes the solution to Eq.~(\ref{eq:OrD}), which provides boundary of the support set $\mathcal{S}$ in the limit of infinitely large $n$.    The eigenvalue outlier is indicated by an arrow in Panel (b).     Panel (a) and (b) show the analytical solution Eq.~(\ref{eq:devCircular}) and Panel (c) is obtained by numerically solving   Eq.~(\ref{eq:OrD}).    }
     \label{fig:example}
 \end{figure*}

 \section{Boundary of  the   spectrum}\label{sec:3x}  
The support set of $\rho(z)$ is defined as
 \begin{equation}
     \mathcal{S} = \overline{\left\{z\in\mathbb{C}: \rho(z)>0\right\}}
 \end{equation} 
 where $\overline{{}\cdot{}}$ denotes the closure of a set.      From  Eq.~(\ref{eq:leadingEigenvalueRho}) it follows that the support set determines the leading eigenvalue whenever the spectrum does not contain eigenvalue outliers \cite{neri2016eigenvalue}, which for the model defined  in Sec.~\ref{sec:2} is the case as  long as $\mu=0$.

The support set  $\mathcal{S}$ follows  from the solutions to   the Eqs.~(\ref{eq:rhoZ})-(\ref{integralResolventEquation}).  The Eq.~\eqref{integralResolventEquation} admits two types of solutions \cite{neri2020linear, mambuca2020dynamical}.   First, there is the trivial solution for which $g_{11} = g_{22} = 0$ and  $\partial_{z^\ast}g_{21} = 0$, yielding  a distribution $\rho=0$ for $z\notin \mathcal{S}$.   Second, there is the nontrivial solution for which 
 $g_{11}>0$ and $g_{22}>0$ and $\partial_{z^\ast}g_{21} \neq 0$, yielding the probability distribution $\rho>0$ for $z\in \mathcal{S}$.   
 
 Although the trivial solution solves the set of Eqs.~(\ref{integralResolventEquation}) for any value of $z$ and for $\eta=0$, it is only for $z\notin \mathcal{S}$ that the trivial solution is relevant.    Indeed,   when $z\in\mathcal{S}$ the trivial solution is unstable with respect to infinitesimal small perturbations, and hence  the regulator $\eta>0$ guarantees that   the spectral distribution  for $z\in\mathcal{S}$ is determined by the nontrivial solution.     As a consequence, the boundary of the support set $\mathcal{S}$ follows  from a linear stability analysis of the Eqs.~(\ref{integralResolventEquation}) around the trivial solution \cite{mambuca2020dynamical}.  Expanding the Eqs.~\eqref{integralResolventEquation} in small values of   $g_{11}>0$ and $g_{22}>0$, we obtain that for all values of $z\in\mathcal{S}$ it holds that 
\begin{equation}
\left\langle \frac{\sigma^2}{\left(D - z^{*}+\tau  \sigma^2 g_{12}\right)\left(D-z+\tau  \sigma^2 g_{21}\right)}\right \rangle_D\geq 1  \label{eq:SupportEq2}
\end{equation}
and the boundary of the support set is given by 
\begin{equation}
\left\langle \frac{\sigma^2}{\left(D - z^{*}+\tau  \sigma^2 g_{12}\right)\left(D-z+\tau  \sigma^2 g_{21}\right)}\right \rangle_D= 1 . \label{eq:SupportEq}
\end{equation}
Note that, in general, Eq.~(\ref{eq:SupportEq}) is coupled with the Eq.~(\ref{integralResolventEquation})
and  therefore these equations have to  be solved together.

In what follows, we first analyse the Eqs.~(\ref{integralResolventEquation})and (\ref{eq:SupportEq}) in two limiting   cases, and then we discuss the general case. 

 \subsection{Symmetric matrices with $J_{ij} = J_{ji}$ ($\tau=1$)} 
 For symmetric random matrices the Eq.~(\ref{integralResolventEquation}) reduces to a functional equation for the resolvent of a Wigner matrix with diagonal disorder derived originally by Pastur in Ref.~\cite{Pastur}.  Indeed, in this case  $g_{22} = g_{11} =0$ for all $z$ with nonzero imaginary part so that  
 \begin{equation}
g_{21} = \int_{\mathbb{R}} {\rm d}x \:p_{D}(x) \frac{1}{z-x-\sigma^2 g_{21}}
 \end{equation}
 for all $z\notin \mathbb{R}$, which  is  identical to  Equation (1.6) in Ref.~\cite{Pastur}.   Since $g_{21}$ is  the Stieltjes transform of the spectral distribution defined on the real line,  we can use the Sokhotski-Plemelj inversion formula (see e.g.~Chapter 8 of \cite{livan2018introduction})
 \begin{equation}
     \rho(x+{\rm i}y) = \frac{1}{\pi} \delta(y) \lim_{\epsilon\rightarrow 0^+}{\rm Im}\left(g_{21}(x-{\rm i}\epsilon)\right) \label{eq:rhoLast}
 \end{equation}
 to obtain the spectral distribution.   Note that the delta distribution $\delta(y)$ on the right hand-side of Eq.~(\ref{eq:rhoLast}) specifies that the eigenvalues of $\mathbf{A}$ are real, and hence the distribution  $\rho$ defined on the complex plane equals  zero for all values $y\neq 0$.

 \subsection{Uncorrelated interaction  variables $J_{ij}$ and $J_{ji}$
($\tau=0$)}  
 In the absence of correlations between $J_{ij}$ and $J_{ji}$, the Eq.~(\ref{eq:SupportEq}) decouples from the Eq.~(\ref{integralResolventEquation}).   Therefore, the "$\tau=0$"-case is mathematically simpler to solve than the "$\tau\neq 0$"-case. The boundary of the support set $\mathcal{S}$ is determined by the values of $\lambda\in\mathbb{C}$ that solve the equation 
 \begin{equation}
    1 =  \sigma^2  \int_{\mathbb{R}} {\rm d}x \; p_D(x) \frac{1}{|\lambda - x|^2}, \label{eq:OrD}
 \end{equation}  
 which is closely related to the results obtained for the boundary of spectra of random directed graphs in Refs.~\cite{neri2016eigenvalue, neri2020linear, tarnowski2020universal} and to those of perturbed random matrices with uncorrelated matrix entries  \cite{khoruzhenko1996large}.

Equation (\ref{eq:OrD}) implies that for $\tau=0$  the leading eigenvalue satisfies
\begin{equation}
  \Re\left( \lambda_1\right) \geq d_{+} = {\rm max}\left\{ x\in\mathbb{R} :  p_D(x)>0\right\}. \label{eq:leadingGaussian}
\end{equation}
In other words, in the absence of correlations between the interaction variables $J_{ij}$ and $J_{ji}$, interactions always  increase the real part of the leading eigenvalue and have thus a destabilising effect on system stability.   

Let us analyse the boundary of the spectrum and the leading eigenvalue for a couple of  examples.   As shown in Appendix~\ref{app:B}, when  $p_D(x)$ is the uniform distribution supported on the interval $[d_-,d_+]$, then the boundary of the support set $\mathcal{S}$ is given by values of $(x,y)$ that solve
\begin{equation}
   \left(\left(d_--x\right) \left(d_+-x\right)+y^2\right) = \frac{y\left(d_+-d_-\right) }{ \tan
   \left( \frac{y}{\sigma^2} \left(d_+-d_-\right)\right)}, \quad  y\in \left(-\frac{\pi\sigma^2}{d_+-d_-}
   ,\frac{\pi\sigma^2}{d_+-d_-}\right)\setminus\left\{0\right\}, \label{eq:devCircular}
\end{equation}
a result that was also obtained in \cite{barabas2017self}.
   For  $(d_+-d_-)/\sigma^2 \ll 1$, we recover the celebrated circular law \cite{girko1985circular, tao2012topics}, while for  $(d_+-d_-)/\sigma^2 \approx 1$ the formula Eq.~(\ref{eq:devCircular})  expresses a deformed circular law replacing the constant radius $\sigma$ by $y\left(d_+-d_-\right)/ \tan
   \left( \frac{y}{\sigma^2} \left(d_+-d_-\right)\right)$.  In Fig.~\ref{fig:example}(a) we have plotted the curve Eq.~(\ref{eq:devCircular}) for the case $d_+=1$ and $d_-=-1$ and we show that this theoretical results is well corroborated by the spectrum obtained from numerically diagonalising a matrix.  From Eq.~(\ref{eq:devCircular})  it follows that the leading eigenvalue is given by 
\begin{equation}
    \Re(\lambda_1)=\frac{1}{2} \left(\sqrt{\left(d_--d_+\right){}^2+4\sigma^2}+d_-+d_+\right). \label{eq:lambda1Analyt}
\end{equation} 
Eq.~(\ref{eq:lambda1Analyt}) reveals that $\Re(\lambda_1)>d_+$ for any value of $\sigma$, and hence the interactions make the system less stable. For $d_+=d_- = d$ we recover the formula Eq.~(\ref{eq:leading1}), and in the limit of large $d_-$ we get $\lim_{d_-\to-\infty}\Re(\lambda_1)=d_+$.

\begin{figure*}[ht]
     \centering
     \subfloat[$\tau=0$]{\includegraphics[scale=0.36]{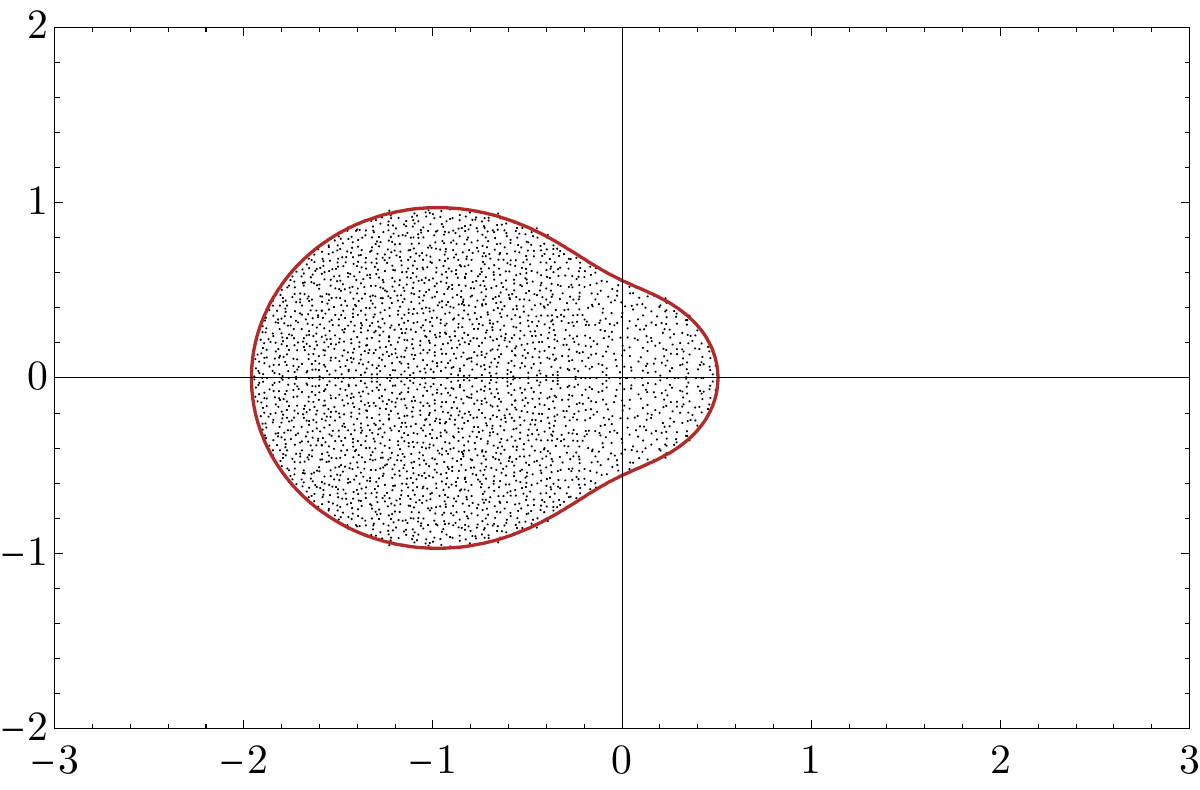}}
     \put(-235,68){$\Im(\lambda)$}
      \put(-190,-12){$\Re(\lambda)$}
     \hspace{0.1mm}
     \subfloat[$\tau=-0.7$]{\includegraphics[scale=0.36]{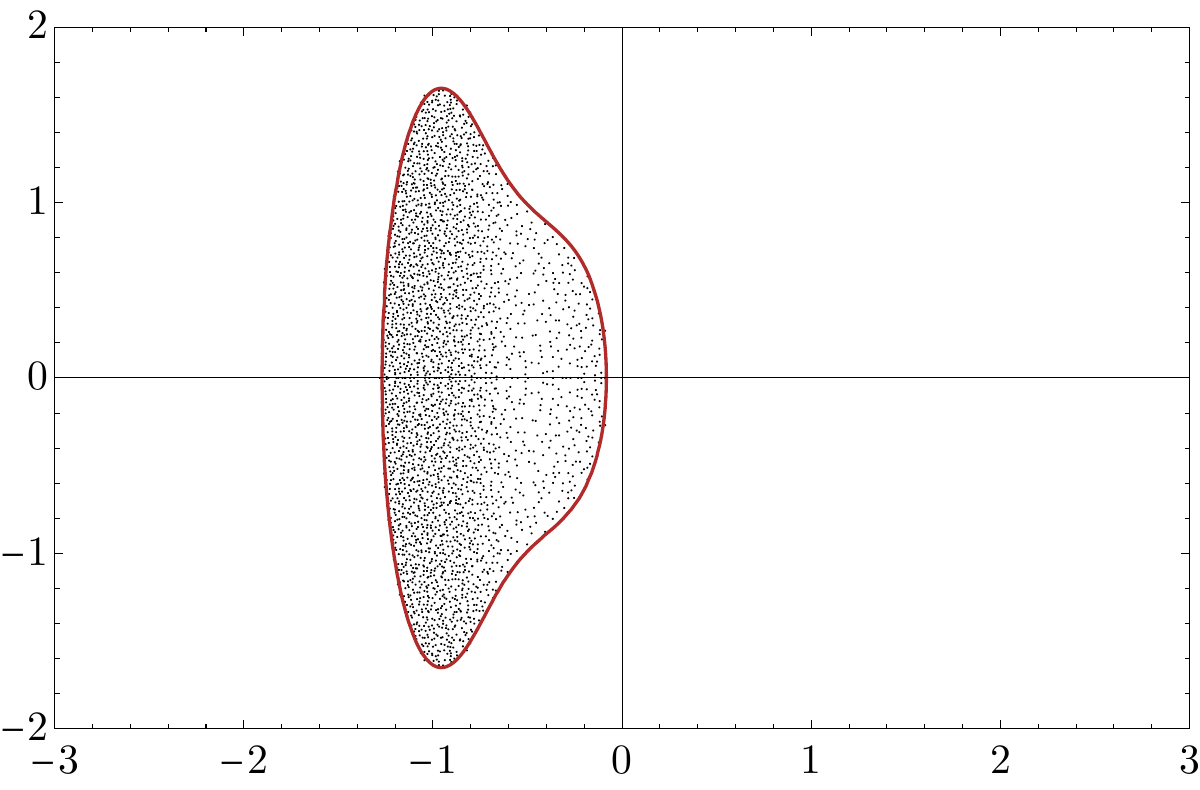}}
     \caption{{\it Comparison between   the spectra  of two random matrices $\mathbf{A}$ with two different values of $\tau$.}  
     Eigenvalues plotted are for two  matrices of size $n=3000$ whose diagonal elements are drawn from the bimodal distribution Eq.~(\ref{eq:bimodalx}) with $d_-=-1$,  $p=0.9$ and $d_+=0.1$, and whose  off-diagonal entries are drawn from a normal distribution with zero mean $\mu=0$, variance $\sigma^2/n = 1/n$, and  $\tau=0$ [Panel (a)] or  $\tau=-0.7$ [Panel (b)].   The red line denotes the solution to the Eqs.~(\ref{re_g12}) and (\ref{eq:boundary_final}).}
     \label{fig:stableRegime}
 \end{figure*}

  As a second example we consider the case when  $p_D$ is a Gaussian distribution with zero mean and unit variance.  In Fig.~\ref{fig:example}(c), we   compare the  solution to  Eq.~(\ref{eq:OrD})  with the spectrum of a random matrix  drawn from the ensemble  defined in Sec.~\ref{sec:2}.  In this case the spectrum $\mathcal{S}$ in the limit $n\rightarrow \infty$ contains the whole real axis, contrarily to the case where $p_D$ is a uniform distribution (compare Fig.~\ref{fig:example}(a) with Fig.~\ref{fig:example}(c)).     The distinction between the two cases follows from the fact that $p_D$ is supported on a compact interval in the uniform case, while it is supported on the whole real axis in the Gaussian case.   Indeed, Eq.~(\ref{eq:OrD})  implies  that  in the former case the spectrum  $\mathcal{S}$ is a finite subset of the complex plane, while in the latter case it contains the real axis.    Consequently, for a compactly supported distribution $p_D$ the leading eigenvalue converges to a finite value as a function of $n$, while for a distribution $p_D$  that is supported on the real axis the leading eigenvalue diverges.   The rate of divergence as a function of $n$ of the average of the leading eigenvalue, $\langle \lambda_1 \rangle$,  is determined by the scaling of the maximum value of the diagonal entries $D_i$ as a function of $n$.    Since the maximum of $n$  i.i.d.~random variables drawn from a Gaussian distribution with zero mean and unit variance scales as $\sqrt{\log(n)}$ (see Theorem 1.5.3 in Ref.~\cite{leadbetter2012extremes}), it holds that
  \begin{equation}
      \langle \lambda_1 \rangle = O_n(\langle D_{\rm max}\rangle) = O(\sqrt{\log (n)})
  \end{equation}
 when $p_D$ is Gaussian,  where $D_{\rm max} = {\rm max}\left\{D_1,D_2,\ldots, D_{n}\right\}$ and where $O(\cdot)$ is the big O notation.
 
\subsection{The case of generic correlations between $J_{ij}$ and $J_{ji}$ ($\tau\in [-1,1]$)} 
We consider now the case of nonzero correlations between the interaction variables $J_{ij}$ and $J_{ji}$.   In this case, it is more difficult to find the values of $z$ that solve the Eq.~(\ref{eq:SupportEq}), as contrarily to the $\tau=0$ case Eq.~(\ref{eq:SupportEq}) is coupled with Eq.~(\ref{integralResolventEquation}).    Nevertheless, we can simplify the Eqs.~(\ref{eq:SupportEq}) and  (\ref{integralResolventEquation}) by using generic properties of $\mathbf{H}$ and $\mathbf{A}$.

Using  that $\mathbf{H}$ is Hermitian, which is   implied by   the definition Eq.~(\ref{eq:14}), we obtain    that  $g_{12}=g^*_{21}$, ${\rm Im}(g_{11}) = 0$ and ${\rm Im}(g_{22}) = 0$.  In addition, since $\mathbf{A}$ and $\mathbf{A}^{T}$ have the  the same statistical properties, we can set $g_{11} = g_{22}$.   Also, since  we are interested in the boundary of the continuous part of the spectrum, which is located at the edge between  the trivial and the nontrivial solutions, we can set  $g_{11}=g_{22}=0$, as this is satisfied for the trivial solution.   Furthermore, we make the ansatz  that $\Im(g_{12})$ is independent of the distribution $p_D$, and therefore   $\Im(g_{12})=y/\sigma^{2}(\tau-1)$, which is the solution when $p_D(x) = \delta(x)$.      In addition, using that $g_{11}g_{22}=0$, we can express the 
 Eq.~(\ref{integralResolventEquation}) as 
 \begin{equation}
    \Re(g_{12}) =\left\langle\frac{D-x+\Re(g_{12})  \tau \sigma^2}{-\left((D-x)+\Re(g_{12}) \tau \sigma^2\right)^2-\frac{y^2}{(1-\tau)^2}}\right\rangle_D
    \label{re_g12}
\end{equation}
and  Eq.~(\ref{eq:SupportEq})  reads
\begin{equation}
    1=\left\langle\frac{\sigma^2}{\left((D-x)+\Re(g_{12}) \tau \sigma^2\right)^2+\frac{y^2}{(1-\tau)^2}}\right\rangle_D.
    \label{eq:boundary_final}
\end{equation} 
We could not simplify these equations further, and hence we will obtain the   boundary of   the spectrum   by solving the Eqs.~(\ref{re_g12}-\ref{eq:boundary_final}).

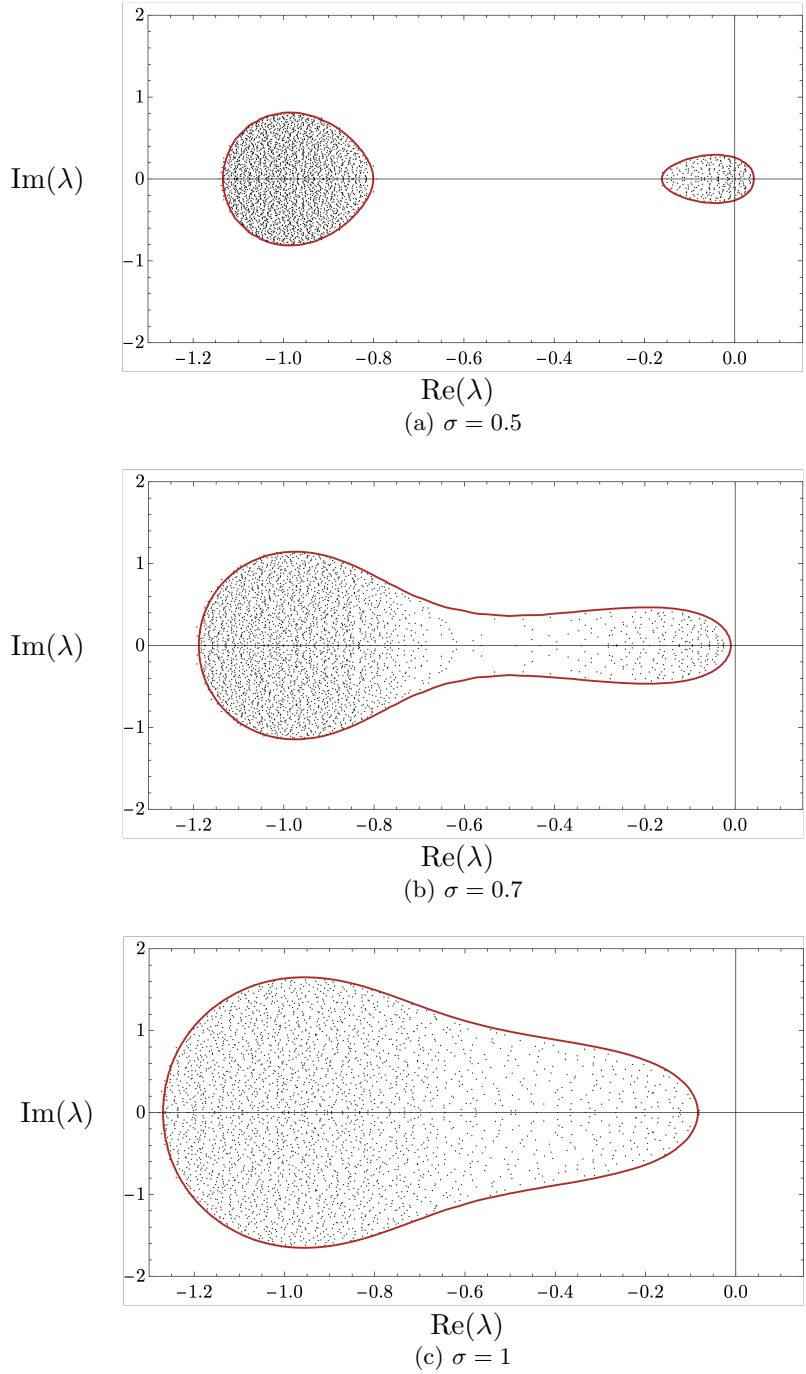
\begin{figure*}[h!]
     \centering
     \subfloat[$\sigma=0.5$]{\includegraphics[scale=0.08]{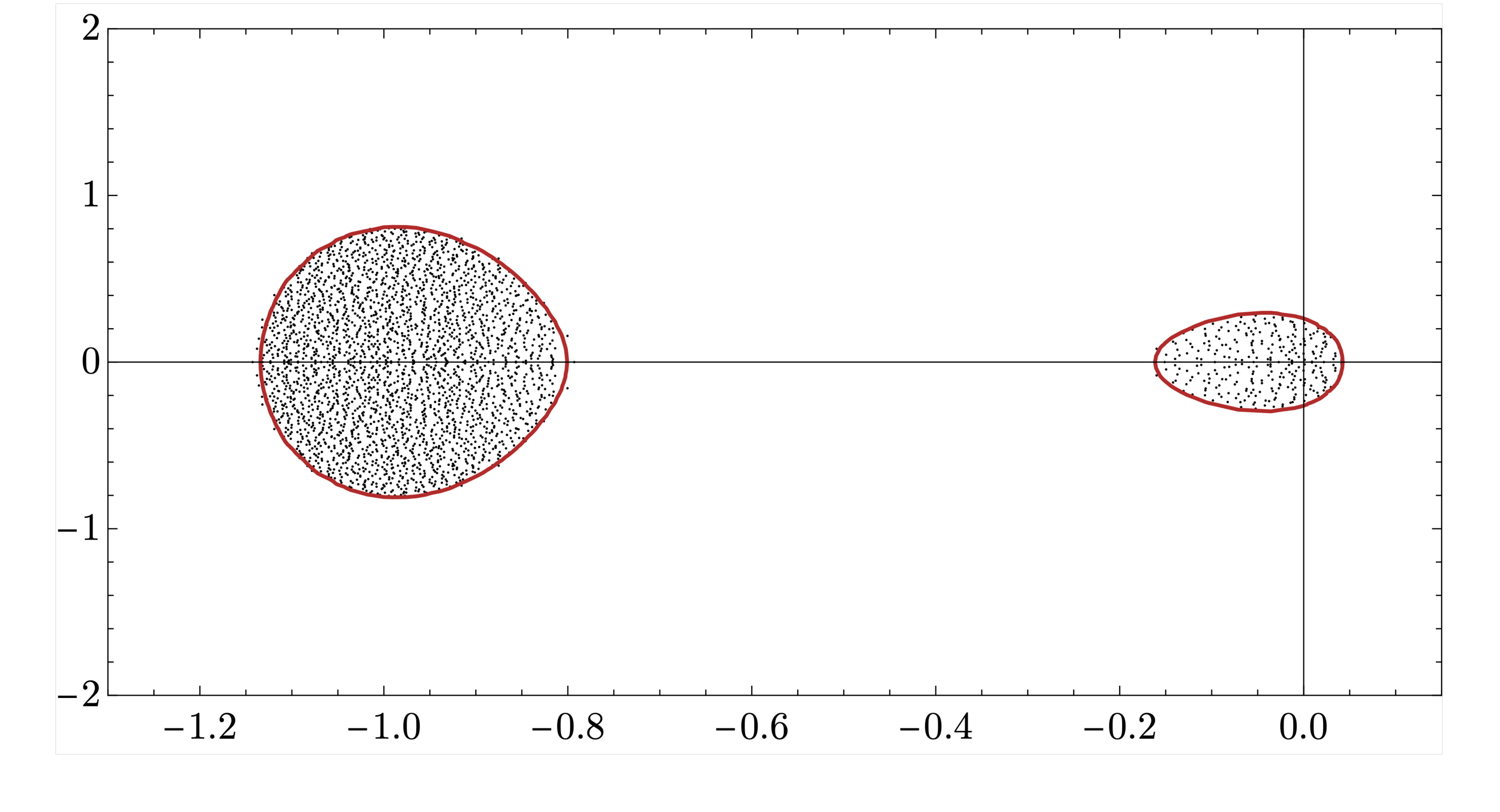}}\label{fig:example2a}
      \put(-310,80){$\Im(\lambda)$}
      \put(-155,0){$\Re(\lambda)$}
    \vspace{0.1mm}
     \subfloat[$\sigma=0.7$]{\includegraphics[scale=0.08]{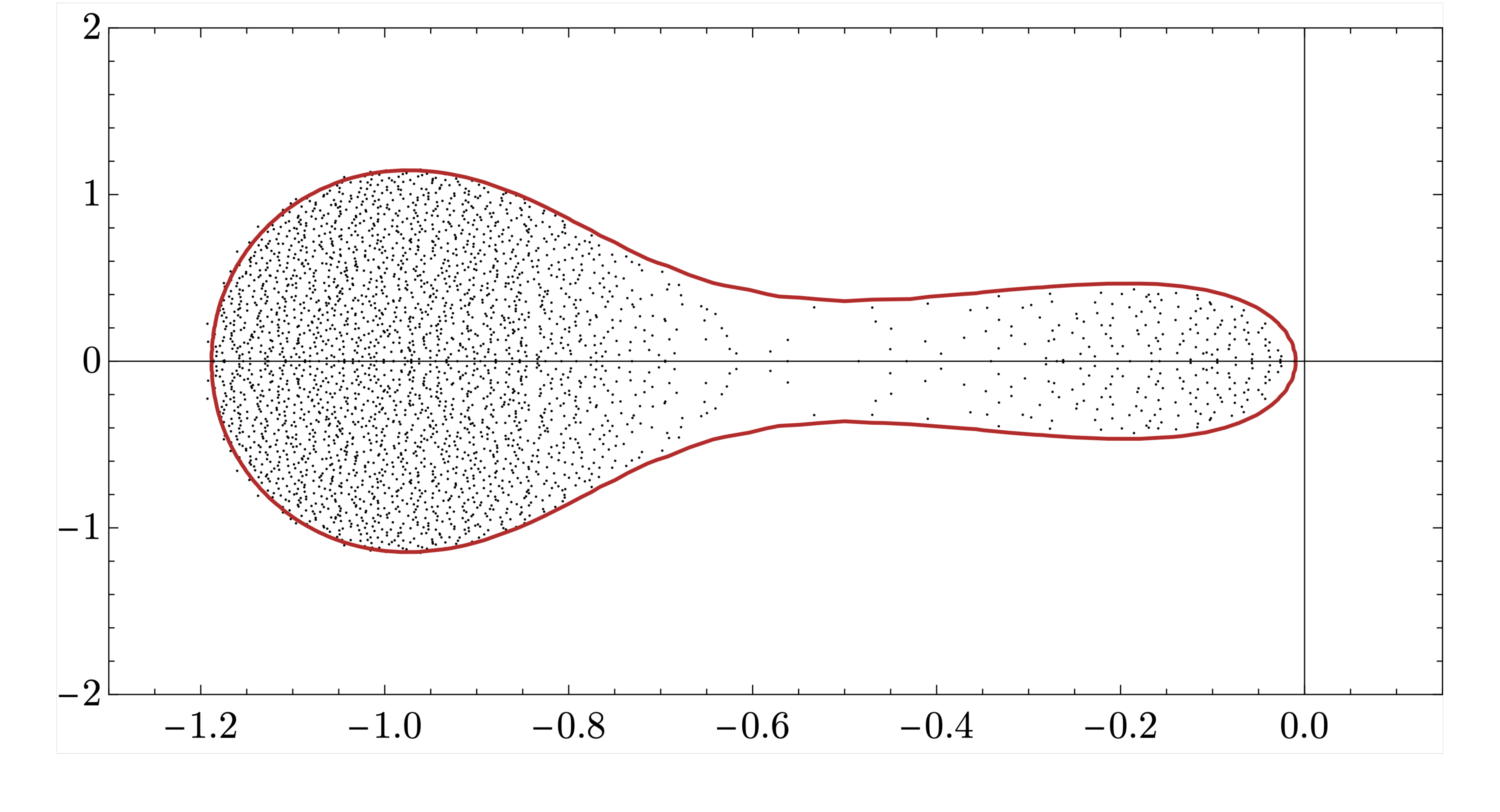}}\label{fig:example2b}
      \put(-310,80){$\Im(\lambda)$}
      \put(-155,0){$\Re(\lambda)$}
           \vspace{0.1mm}
    \subfloat[$\sigma=1$]{
    \includegraphics[scale=0.08]{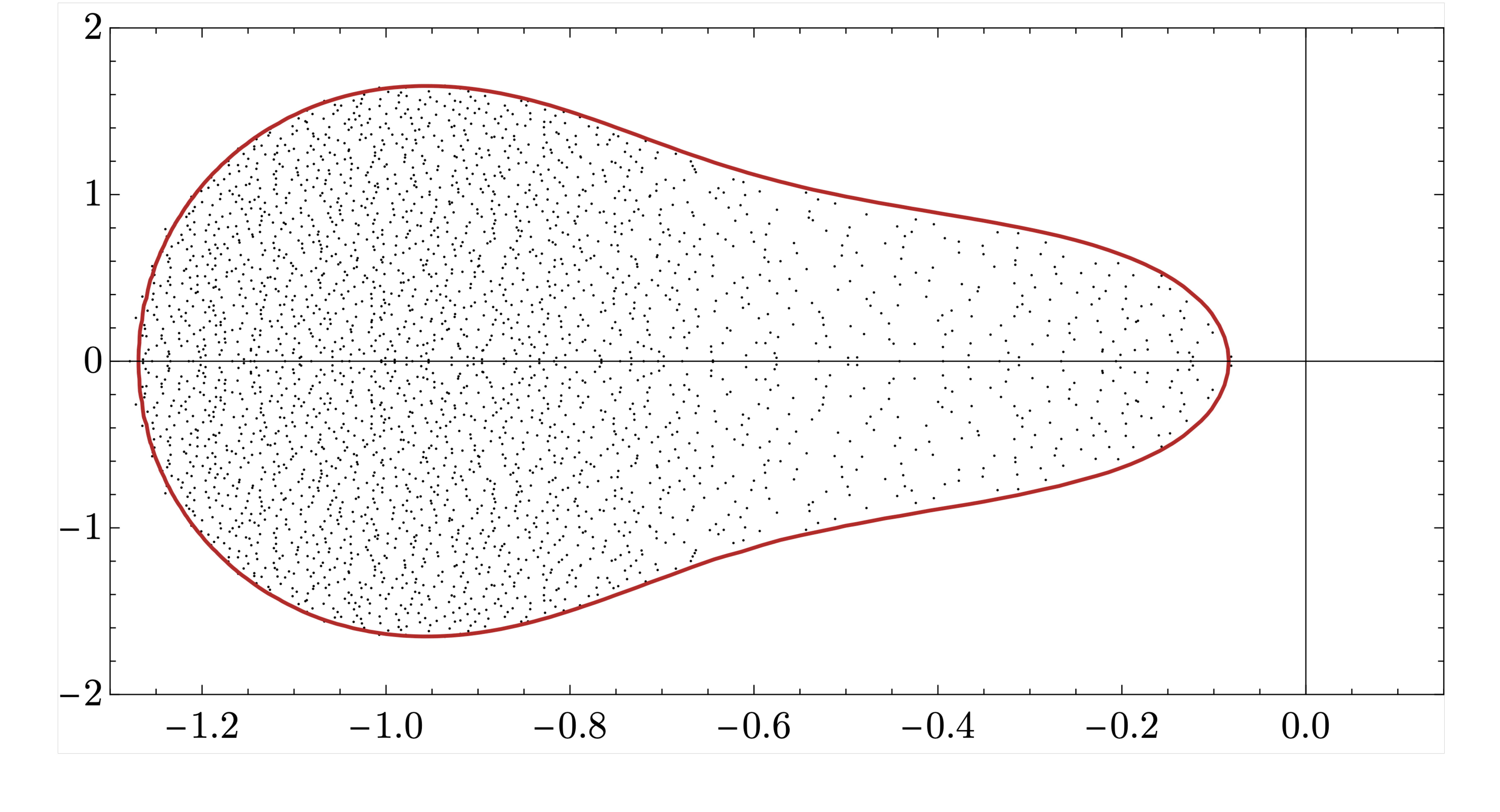}
    }
     \label{fig:example2c}
      \put(-310,80){$\Im(\lambda)$}
      \put(-155,0){$\Re(\lambda)$}
     \caption{{\it Comparison between the spectra of random matrices $\mathbf{A}$ with different values of the interaction strength $\sigma$.} Eigenvalues plotted are for three matrices  of size $n=3000$ whose  off-diagonal elements $(J_{ij},J_{ji})$  are drawn from a joint Gaussian distribution with zero mean,  a Pearson correlation coefficient $\tau=-0.7$, and a variance $\sigma^2/n$ as indicated.    The diagonal elements follow a bimodal distribution with parameters $p=0.9$, $d_-=-1,d_+=0.1$, and $\mu=0$. The red line denotes the solution to the Eqs.~(\ref{re_g12}) and (\ref{eq:boundary_final}). }
     \label{fig:exampleRetraction}
 \end{figure*}

In Figs.~\ref{fig:stableRegime} and \ref{fig:exampleRetraction}, we corroborate the boundary of the spectrum, obtained from solving the  Eqs.~(\ref{re_g12}-\ref{eq:boundary_final}), with numerical results for the eigenvalues of matrices of finite size, obtained with  numerical diagonalisation routines.   We show the boundary of the spectrum for the case of the bimodal distribution $p_D$ given by Eq.~(\ref{eq:bimodalx}).  Figure~\ref{fig:stableRegime} compares two spectra with the same $\sigma$ but different values of $\tau$, whereas Fig.~\ref{fig:exampleRetraction} considers one negative value of $\tau$ and observes how the spectrum changes as a function of $\sigma$.   Note that the  the real part of the  leading eigenvalue $\Re(\lambda_1)$ decreases as a function of $\tau$.

The leading eigenvalue is obtained by solving Eqs.~(\ref{re_g12})-(\ref{eq:boundary_final}) at $y=0$.   For bimodal $p_D$ we obtain a quartic equation in $x$ and we identify the largest real-valued solution of this quartic equation with  ${\rm Re}(\lambda_1)$.  We have obtained an analytical expression for ${\rm Re}(\lambda_1)$ as a function of the system parameters, which we omit here as it is a  long mathematical formula without clear use.  However, it can be found in the Supplemental Material of  Ref.~\cite{barabas2017self}.      

For uniform $p_D$ the Eqs.~(\ref{re_g12})-(\ref{eq:boundary_final})  can  be solved explicitly as shown in Appendix~\ref{AppD}.  Remarkably, in this case we obtain a simple, analytical expression for  the leading eigenvalue, viz., 
 \begin{eqnarray}
\text{Re}(\lambda_1)&=&\frac{1}{2} \left(\sqrt{\left(d_--d_+\right){}^2+4 \sigma
   ^2}+d_-+d_+\right)  \nonumber\\ 
  && +\tau \frac{\sigma ^2 }{d_+-d_-} \log
   \left(\frac{\sqrt{\left(d_--d_+\right){}^2+4 \sigma
   ^2}+d_+ -d_-}{\sqrt{\left(d_--d_+\right){}^2+4 \sigma
   ^2}-d_++d_-}\right),  \label{eq:lambda1Analyt2}
\end{eqnarray} 
where $d_+>d_-\in \mathbb{R}$.  
One readily verifies that for $\tau=0$ Eq.~(\ref{eq:lambda1Analyt2}) reduces to Eq.~(\ref{eq:lambda1Analyt}),  for $\tau=1$ Eq.~(\ref{eq:lambda1Analyt2}) recovers the result in Ref.~\cite{mergny2021stability}  for the case of symmetric matrices with entries drawn from a Gaussian distribution, and for $\sigma=1$ it is equivalent to a formula that appeared in the Supplemental Material of \cite{barabas2017self}.   Since the sign of the second term of Eq.~(\ref{eq:lambda1Analyt2})  is equal to the sign of $\tau$,  the leading eigenvalue $\lambda_1$ decreases as a function of negative values of  $\tau$.

\begin{figure}[h!]
     \centering
     \includegraphics[scale=0.72]{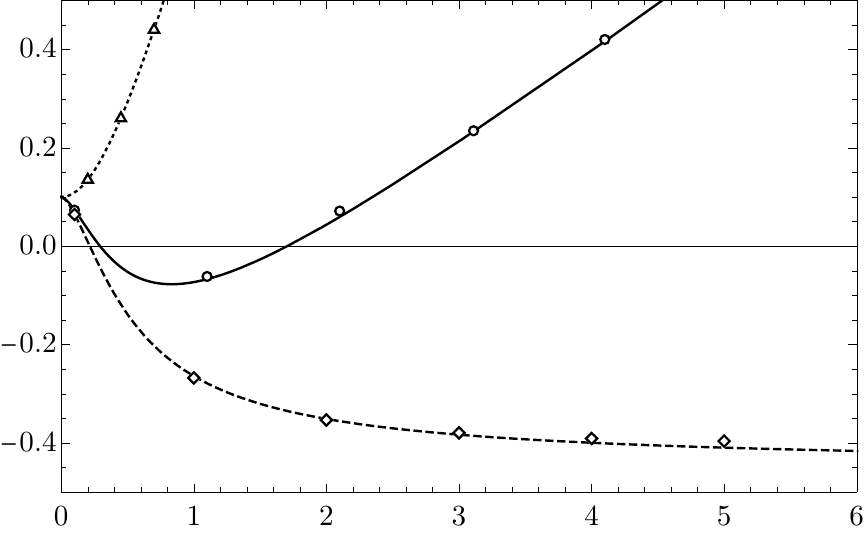}
     \put(-335,98){$\Re(\lambda_1)$}
      \put(-140,-9){$\sigma$}
     \caption{{\it Effect of the  interaction strength $\sigma$ on the real part of the leading eigenvalue $\lambda_1$ when $\mu=0$ and  for  $\tau = 0$ (triangle, dotted), $\tau=-0.8$ (circle, solid) and $\tau=-1$ (diamond, dashed).}   Lines show the Eq.~(\ref{eq:lambda1Analyt2}).  Markers are numerical results obtained  for random matrices $\mathbf{A}$ with diagonal elements $D_j$ that are  drawn independently from a uniform $p_D$ supported on the  interval  $ [d_-,d_+] = [-1,0.1]$ and with  pairs of off-diagonal elements $(J_{ij},J_{ji})$ that are drawn independently  from a normal distribution with mean $0$, variance $\sigma^2/n$, and Pearson correlation coefficient $\tau$ as provided.     Each marker represents the largest eigenvalue of one matrix realisation of size $n=7000$.  }
     \label{fig:reEntranceEffect}
\end{figure}

   In Fig.~\ref{fig:reEntranceEffect} we compare Eq.~(\ref{eq:lambda1Analyt2})  with numerical results of the leading eigenvalue obtained through the direct diagonalisation of matrices of finite size $n$.   The numerics corroborate well the analytical results that are valid for infinitely large $n$.  We make a few interesting observations from Fig.~\ref{fig:reEntranceEffect}:  (i) for $\tau=0$, the leading eigenvalue is a monotonically increasing function of  the interaction  strength $\sigma$ implying a continuous increase of the width of the spectrum as a function of $\sigma$; (ii)  for $\tau=-0.8$, the leading eigenvalue is a nonmonotonic function of $\sigma$.  Initially, for small values of $\sigma$, the width of the spectrum decreases as a function of $\sigma$, while for large enough values of $\sigma$ the width of the spectrum increases as a function of $\sigma$; (iii) for $\tau=-1$, the leading eigenvalue is monotonically decreasing.   In this case, the width of the spectrum decreases continuously as a function of $\sigma$ and converges for large $\sigma$ to a vertical spectrum centered on the mean value of $p_D$.

\section{Eigenvalue outlier}\label{sec:outlier}
Now, we determine the leading eigenvalue when $\mu\neq0$.   Even though, the continuous part of the spectrum is not affected by $\mu$ (see Appendix~\ref{app:muDiffZero}), the  spectrum may have an eigenvalue outlier, which can be the leading eigenvalue; this is illustrated in Panel (b) of Fig.~\ref{fig:example}.    Hence, for $\mu\neq 0$ the leading eigenvalue 
\begin{equation}
    \lambda_1 = {\rm max}\left\{\lambda^{\rm c}_1,\lambda_{\rm isol}\right\} \label{eq:lambda1F}
\end{equation}
where $\lambda_{\rm isol}$ is the eigenvalue outlier, if it exists, and $\lambda^{\rm c}_1$ is the leading eigenvalue of the continuous part of the spectrum, as defined by Eq.~(\ref{eq:leadingEigenvalueRho}) with $\lambda_1$ replaced by $\lambda^{\rm c}_1$.   In what follows, we determine $\lambda^{\rm c}_1$ and $\lambda_{\rm isol}$.

The leading eigenvalue $\lambda^{c}_1$ of the support set $\mathcal{S}$ is, in the limit of large $n$, independent of $\mu$.   Indeed,  as shown in Appendix~\ref{app:muDiffZero}, the boundary of the support set $\mathcal{S}$ solves the Eqs.~(\ref{integralResolventEquation}) and (\ref{eq:SupportEq}), just as was the case for $\mu=0$.     

To determine the eigenvalue outlier we follow the theory for eigenvalue outliers of random matrices, as developed in  Ref.~\cite{neri2020linear}, which is also based on the cavity method, albeit works in a different way as in this approach recursion relations are derived for entries of right eigenvectors, instead of the entries of the resolvent.   Following this approach, we show  in the Appendix~\ref{app:outlier2} that the eigenvalue outliers $\lambda_{\rm isol}$ of $\mathbf{A}$ in the limit for large $n$ solve
\begin{equation}
    1 = \mu \: g_{21}(\lambda_{\rm isol}), \quad \lambda_{\rm isol}\notin\mathcal{S},
    \label{eq_outlierx}
\end{equation} 
where $g_{21}$ is the  trivial solution to  Eq.~(\ref{integralResolventEquation}), i.e., for  $g_{11}=g_{22}=0$.    For $\tau=0$,
\begin{equation}
    g_{21}(z) = \Big\langle \frac{1}{z - D} \Big\rangle_D,
\end{equation}
which leads to the equation 
\begin{equation}
1 = \mu  \Big\langle \frac{1}{\lambda_{\rm isol} - D} \Big\rangle_D
\end{equation}
for the outlier $\lambda_{\rm isol}$.  

In general for $\tau\neq 0$, we do not have an explicit expression for $g_{21}$, and hence we express $\lambda_{\rm isol}$  in terms of the functional inverse $f_{21}$ of $g_{21}$, namely, 
\begin{equation}
  \lambda_{\rm isol} =   f_{21}\left(\frac{1}{ \mu }\right) ,
    \label{eq_outlier_inverse}
\end{equation} 
where
\begin{equation}
z = f_{21}(g_{21}(z)) .    
\end{equation}

Using Eq.~(\ref{integralResolventEquation}) for $g_{22}= g_{11} = 0$,  corresponding to the trivial solution, we obtain the following selfconsistent equation for $g_{21}(z)$, 
\begin{equation} 
g_{21} =\Big\langle  \frac{1}{z-D-\tau\sigma^2g_{21}}\Big\rangle_{D},  \quad  {\rm with} \quad  z\notin \mathcal{S}.
\label{trivial_eq_resolvent}
\end{equation}    
If we can rewrite Eq.~(\ref{trivial_eq_resolvent})  as  $z = f_{21}(g_{21})$, then we readily obtain the functional inverse $f_{21}$ of $g_{21}$.

Now, we determine $f_{21}$ for specific distributions of the diagonal disorder $D$.  First, we consider the case when $p_D$ is uniform, as in Eq.~(\ref{eq:uniform}).   It then holds     for real values of $z\in \mathbb{R}$ that $\text{Im}(g_{21})=0$, and 
\begin{eqnarray}
\text{Re}(g_{21}) &=&\frac{1}{d_+-d_-} \int^{d_+}_{d_-}{\rm d}u \frac{1}{z-u-\tau\sigma^2\text{Re}(g_{21})} 
\nonumber\\ 
&=& \frac{1}{d_+-d_-} \log \frac{z-d_--\tau\sigma^2\text{Re}(g_{21})}{z-d_+-\tau\sigma^2\text{Re}(g_{21})},  \label{eq:zUniformG2}
\end{eqnarray}
from which it follows that 
\begin{eqnarray}
z & = & d_-+ \frac{d_-+d_+}{e^{-(d_p-d_m)\text{Re}(g_{21})}}+\tau\sigma^2\text{Re}(g_{21}) ,\label{eq:zUniformG}
\end{eqnarray}
and thus 
\begin{equation}
    f_{21}(u) = d_-+ \frac{d_-+d_+}{e^{-(d_p-d_m)\text{Re}(u)}}+\tau\sigma^2\text{Re}(u). \label{eq:f21Uniform}
\end{equation} 
Inserting the expression  Eq.~(\ref{eq:f21Uniform}) into Eq.~(\ref{eq_outlier_inverse}), we obtain 
\begin{equation}
    \lambda_{\rm isol} =  d_-+ \frac{d_--d_+}{e^{-\frac{(d_+-d_-)}{\mu }}-1}+\tau\frac{\sigma^2}{\mu},
    \label{outlier_uniform}
    \end{equation}
which is an explicit analytical expression for the outlier when $D$ is uniformly distributed.     

\begin{figure}[h!]
     \centering
     \includegraphics[scale=0.72]{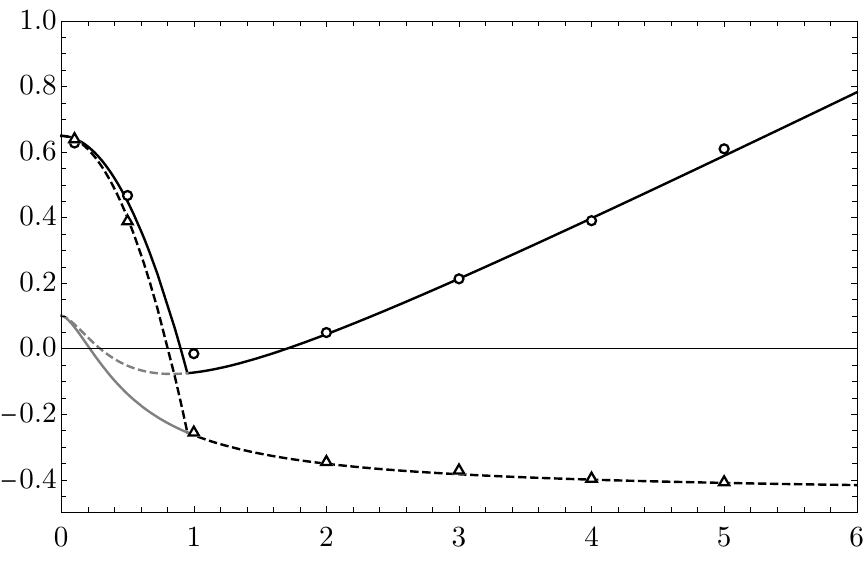}
      \put(-335,98){$\Re(\lambda_1)$}
      \put(-145,-9){$\sigma$}
     \caption{{\it Effect of the  interaction strength $\sigma$ on the real part of the leading eigenvalue $\lambda_1$ for random matrices with  $\mu=2$ and all other parameters the same as in  Fig.~\ref{fig:reEntranceEffect}.}   Similar to Fig.~\ref{fig:reEntranceEffect}, solid lines/circles correspond  with $\tau=-0.8$ and dashed lines/triangles with $\tau=-1$.  Gray lines show Eq.~(\ref{eq:lambda1Analyt2}). Black lines show the maximum between Eq.~(\ref{outlier_uniform}), for the eigenvalue outlier, and  Eq.~(\ref{eq:lambda1Analyt2}), for the leading eigenvalue of $\mathcal{S}$. Each marker represents the largest eigenvalue of one matrix realisation of size $n=3000$.  }
     \label{fig:reEntranceEffect_with_outlier}
\end{figure}

Equation~(\ref{outlier_uniform}) shows that for negative $\tau$ the eigenvalue outlier  decreases monotonically as a function of $\sigma$, which is different from the nonmonotonic behaviour  of $\lambda^{\rm c}_{1}$ as a function of $\sigma$, see Eq.~(\ref{eq:lambda1Analyt2}).  
In Fig.~\ref{fig:reEntranceEffect_with_outlier}, we plot $\lambda_{1} = {\rm max}\left\{\lambda^{\rm c}_{1},\lambda_{\rm isol}\right\}$ as a function of $\sigma$.   For small values of $\sigma$, the leading eigenvalue $\lambda_1$ decreases rapidly as a function of $\sigma$, as $\lambda_1$ is an outlier, until $\lambda_{\rm isol}=\lambda^{\rm c}_{1}$, at which point the outlier stops existing and $\lambda_1$ is located at the boundary of $\mathcal{S}$.    

Also for the case of bimodal $p_D$, as given by Eq.~(\ref{eq:bimodalx}), we can obtain an explicit expression for $\lambda_{\rm isol}$.    Following the same steps as for uniform disorder, we get 
\begin{equation}
\lambda_{\rm isol} = \frac{1}{2}\left(d_-+d_+ +\mu +\sqrt{( d_+-d_-)^2+2(d_--d_+)(2p-1)\mu+\mu^2 }\right) +\tau \frac{\sigma^2}{\mu }.
\label{outlier_bimodal}
\end{equation}
Again, as in the case for uniform disorder, the outlier decreases monotonically as a function of $\sigma$ when $\tau<0$.    

Comparing Eqs.~(\ref{outlier_uniform}) with (\ref{outlier_bimodal}), we make the following interesting observation.   Both equations take the form 
\begin{equation}
\lambda_{\rm isol} = \lambda^{(0)}_{\rm isol} + \tau \frac{\sigma^2}{\mu}   \label{eq:general}   
\end{equation}
where $\lambda^{(0)}_{\rm isol}$ is the  corresponding eigenvalue outlier for $\tau=0$ solving
\begin{equation}
1 = \mu  \Big\langle \frac{1}{\lambda^{(0)}_{\rm isol} - D} \Big\rangle_D.   \label{eq:pDTauZero}
\end{equation}
Since Eq.~(\ref{eq:general}) holds for both uniform and bimodal $p_D$, we conjecture that  Eq.~(\ref{eq:general}) holds for general $p_D$.   The Eq.~(\ref{eq:general})  is a convenient result as  $\lambda^{(0)}_{\rm isol}$ can be obtained easily from solving  Eq.~(\ref{eq:pDTauZero}).     Notice that for constant diagonal matrix entries, i.e., $D_i=d$ for all $i\in [n]$,  Eq.~(\ref{eq:general}) is consistent with Theorem 2.4 of Ref.~\cite{o2014low} for the eigenvalue outliers of  finite rank perturbations of elliptic random matrices.

\section{Stability of linear dynamical systems}\label{sec:4} 
We discuss the implications of the spectral results obtained in the previous two sections for  the  stability of  linear systems of the form given by Eq.~(\ref{eq:1}).  

\subsection{Uncorrelated interactions destabilise dynamical systems} 
For $\tau=0$ it holds that $\Re(\lambda_1)\geq d_+$ for all values of $\sigma$ [see Eq.~(\ref{eq:leadingGaussian})], which has a couple of interesting implications for the stability of linear dynamical systems.   First, a linear dynamical system with  $\tau=0$  cannot be stable if the support of $p_D$ covers the positive axis.  Second,   interactions $J_{ij}$  destabilise  linear dynamical systems as $\lambda_1$ is an increasing function of $\sigma$ (see also Fig.\ref{fig:exampleRetraction}).     Third, if the support of $p_D$ covers the whole real line, then the leading eigenvalue $\lambda_1$ diverges as a function of $n$.   In the latter case we obtain a tradeoff between diversity, as measured by $n$, and stability, as measured by $\Re(\lambda_1)$~\cite{gardner1970connectance, may1972will}.  Indeed, when $p_D$ has unbounded support, then for any realisation of the system parameters $\sigma$, $\tau$, and $p_D$, there will exist a value $n^\ast$ so that with large probability $\Re(\lambda_1)>0$ when $n>n^\ast$.   In Ref.~\cite{mambuca2020dynamical}, the latter scenario is  referred to as size-dependent stability, as the system size $n$ is  an important parameter in determining  system stability.  

\subsection{Antagonistic interactions can render dynamical systems stable}
In the case of negative $\tau$ values the interactions $J_{ij}$ can stabilise linear dynamical systems when they are neither too strong nor too weak.  To understand how this works, consider linear systems $\mathbf{A}$ for which there exist   values $x\in\mathbb{R}^+$ with $p_D(x)>0$, such that the system is unstable in the absence of interactions.  As illustrated in Fig.~\ref{fig:exampleRetraction}, adding antagonistic interactions  to  a linear system can retract the real part $\Re(\lambda_1)$ of the leading eigenvalue  and make it  negative. This example demonstrates that unlike the uncorrelated case with $\tau=0$, interactions can contribute to the stability of a system when $\tau<0$.   However, as shown in Figs.~\ref{fig:reEntranceEffect} and \ref{fig:reEntranceEffect_with_outlier}, for  large  values of  the interaction strength $\sigma$ the leading eigenvalue increases as a function of $\sigma$, and hence antagonistic interactions stabilise linear dynamical systems as long as they are neither too strong nor too weak.

\begin{figure}
     \centering
     \includegraphics[width=0.618\textwidth]{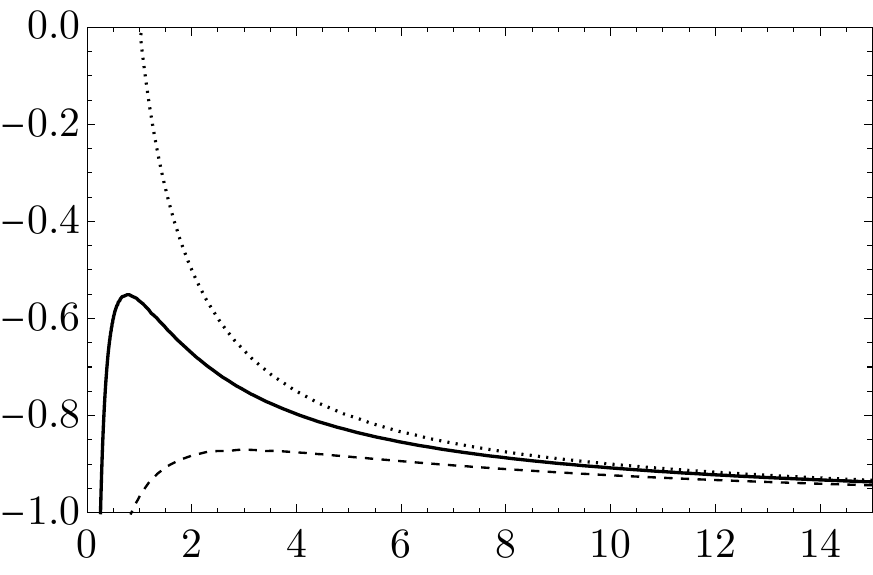}
    \put(-285,90){\Large $\tau$}
      \put(-120,-12){\Large$\sigma$}
     \caption{{\it Phase diagram for the stability of  linear dynamical systems with antagonistic interactions when $\mu\leq 0$.}  Lines denote values of $(\tau,\sigma)$  of marginal stability, i.e.   $\Re(\lambda_1)=0$, separating  a stable region with $\Re(\lambda_1) < 0$ (below the lines) from an unstable region with   $\Re(\lambda_1) >0$ (above the lines).  Results shown are for the random matrix model defined in Sec.~\ref{sec:2} and for various distributions $p_D$ with the same mean  $\langle D\rangle = -1$.    The solid line represents a uniform disorder on the interval $I= [-2.1,0.1]$; the dashed line represents a bimodal disorder with parameters $p=0.5$, $d_-=-2.1,d_+=0.1$; and the dotted line represents the case where all diagonal elements take the value $-1$ with no disorder.
    }
     \label{fig:phaseDiagCorrelation}
 \end{figure}

Fig.~\ref{fig:phaseDiagCorrelation} draws the lines of  marginal stability, corresponding with  ${\rm Re}(\lambda_1)=0$, in the $(\sigma, \tau)$ plane for $\mu\leq 0$ and  for homogeneous growth rates $p_D(x) = \delta(x-D)$ (dotted line),  for a bimodal distributions $p_D$ (dashed line), and for  a uniform distribution $p_D$ (solid line).  In these cases, the leading eigenvalue $\lambda_1$ is located at the boundary of the support set $\mathcal{S}$, such that, $\lambda_1=\lambda^{\rm c}_1$.  For all cases we have set  $\langle D\rangle = -1$, so that we see the effect of fluctuations in $D$ on system stability.    Note that for the dotted line $d_+=-1$, whereas for the dashed and solid lines $d_+=0.1$.   As a consequence, for the dotted line a  stable region exists  when $\tau=0$ and $\sigma$ is small enough, while for the other cases there is no stable region when $\tau=0$.      Interestingly, for negative values of  $\tau$ and for interaction strengths $\sigma$ that are neither too weak nor too strong, there  exists  a stable region with  $\Re(\lambda_1)<0$.     This   region exists even though $d_+>0$ (solid and dashed lines).     On the other hand, for $\tau=0$ a stable region can only exist when $d_+<0$, which is the case of the dotted line with homogeneous rates.  
 
 Figure~\ref{fig:phaseDiagCorrelation} shows that $\Re(\lambda_1)$ is for fixed $\langle D\rangle$ and large $\sigma$ independent of $p_D$.  We  explore this universal behaviour in more depth.   
 Expanding the expression of $\Re(\lambda_1)$ for Eq.~\eqref{eq:lambda1Analyt2} in large values of $\sigma$ we obtain
\begin{equation}
 \Re(\lambda_1)=(1+\tau )\sigma +\frac{1}{2} \left(d_-+d_+\right)+(3+\tau)\frac{
   \left(d_--d_+\right){}^2}{24 \sigma}+O(1/\sigma^2).  \label{eq:RelargeSigma}
\end{equation}
Identifying the mean and  variance of the uniform distribution $p_D$ in Eq.~(\ref{eq:RelargeSigma}), we can write
\begin{equation}
    \Re(\lambda_1)= (1+\tau)\sigma+\langle D\rangle -   \langle\langle D^2\rangle\rangle \frac{(\tau-3)}{2\sigma} +O(1/\sigma^2), \label{eq:leadingLargeSigma}
\end{equation}
where $\langle\langle D^2\rangle\rangle$ represents the variance of the diagonal elements.   If $D$ is a deterministic variable with zero variance, then we recover the Eq.~\eqref{eq:leading1}. This suggests that when the interactions are strong enough, only the first moment of the diagonal elements is important, rather than the distribution of their elements. Although the relation Eq.~(\ref{eq:leadingLargeSigma}) is  derived for the uniform case, numerical evidence shows that it also holds for the bimodal case, and therefore we conjecture that it holds for arbitrary $p_D$ distributions.   Demonstrating the validity of  the Eq.~(\ref{eq:leadingLargeSigma}) beyond the uniform case would be an interesting  extension of the present work.

 \begin{figure}
     \centering
     \centering
     \subfloat[$\mu=1$]{\includegraphics[width=0.49\textwidth]{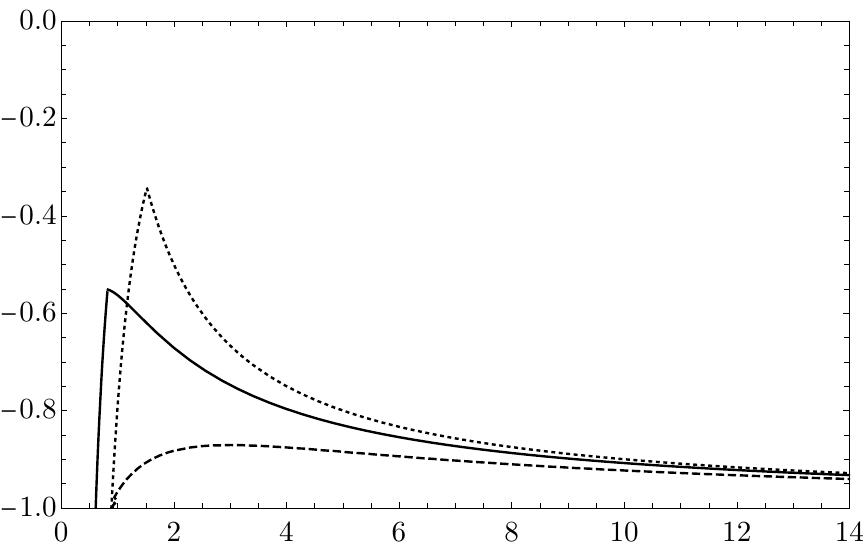}}
      \hspace{1mm}
      \subfloat[$\mu = 5$]{\includegraphics[width=0.49\textwidth]{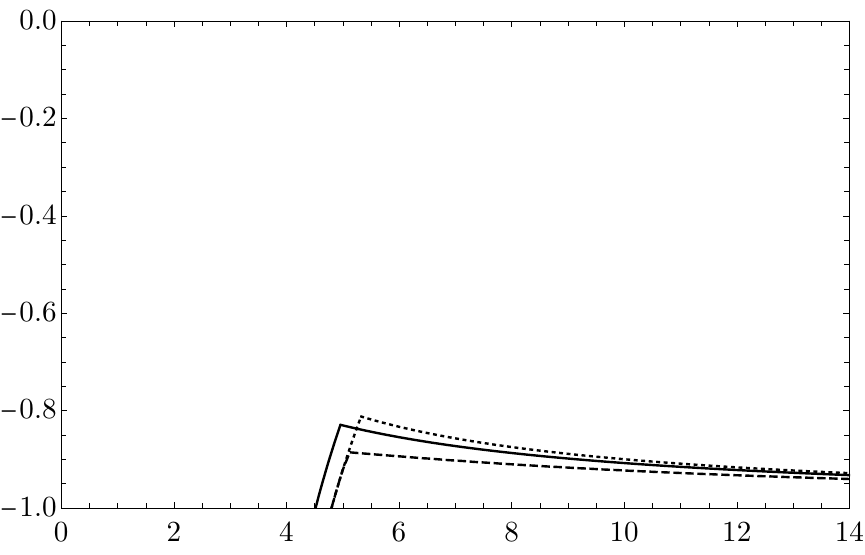}}
        \put(-438,68){ $\tau$}
      \put(-317,1){$\sigma$}
      \put(-220,68){ $\tau$}
      \put(-103,1){$\sigma$}
     \caption{{\it Phase diagram for the stability of  linear dynamical systems with antagonistic interactions  and for  $\mu> 0$.}  Lines show parameter values for which the system is marginally stable ($\Re[\lambda_1]=0$).    The parameters are the same as in Fig.~\ref{fig:phaseDiagCorrelation}, except for $\mu$, which is set to  $\mu = 1$ in Panel (a) and $\mu= 5$ in Panel (b).
    }
     \label{fig:phaseDiagCorrelation_with_outlier}
 \end{figure}

Figure~\ref{fig:phaseDiagCorrelation_with_outlier} draws the lines of  marginal stability in the $(\sigma, \tau)$, similar to Fig.~\ref{fig:phaseDiagCorrelation}, albeit with  $\mu>0$.    In this case, the leading eigenvalue $\lambda_1$ is  either the eigenvalue outlier, $\lambda_1 = \lambda_{\rm isol}$, when it exists, or is the leading eigenvalue of the support set $\mathcal{S}$, i.e., $\lambda_1=\lambda^{\rm c}_1$.    The eigenvalue outlier exists when $\sigma$ is small enough, and in this regime the system stability increases as a function of $\sigma$.  For fixed $\tau$, at a value $\sigma=\sigma^\ast$ the eigenvalue outlier merges into $\mathcal{S}$, i.e., $\lambda_{\rm isol} = \lambda^{\rm c}_1$, and for $\sigma>\sigma^\ast$ the leading eigenvalue is $\lambda^{\rm c}_1$.    We can clearly  notice this transition in Fig.~\ref{fig:phaseDiagCorrelation_with_outlier} due to the cusp that appears in the lines of marginal stability.  Hence, for $\sigma>\sigma^\ast$, the lines of marginal stability in Fig.~\ref{fig:phaseDiagCorrelation_with_outlier}  are identical to those in Fig.~\ref{fig:phaseDiagCorrelation}. Comparing Panels (a) and (b) in  Fig.~\ref{fig:phaseDiagCorrelation_with_outlier}, we notice that increasing the   parameter $\mu$ reduces system stability.   This can be understood from the fact that the eigenvalue outlier $\lambda_{\rm isol}$ increases as a function of $\mu$, while the boundary of the continuous spectrum is independent of $\mu$.   Note that negative values of $\mu$ have no influence on system stability as in this case the eigenvalue outlier will be located on  real axis at the negative side of $\mathcal{S}$, when it exists.

\section{Discussion}\label{sec:5}
We have obtained exact results for the leading eigenvalue  of random matrices of the form Eq.~(\ref{eq:AD}), where the pairs $(J_{ij},J_{ji})$ are i.i.d.~random variables drawn from a joint distribution with moments as given in Eq.~(\ref{momentsOfMatrixJ}), and where the diagonal elements $D_i$ are i.i.d.~random variables drawn from a distribution $p_D$.   

If the Pearson correlation coefficient $\tau=0$, then the boundary of the spectrum solves the  Eq.~(\ref{eq:OrD}), which implies that  $\lambda_1\geq d_+$ [see Eq.~(\ref{eq:leadingGaussian})].  Hence, in this case interactions render dynamical systems less stable,  irrespective of the form of $p_{J_1,J_2}$ and $p_D$.   On the other hand, if the Pearson correlation coefficient  $\tau$ between the pairs $(J_{ij},J_{ji})$ is negative and the variance of the distribution $p_D$ is nonzero, then $\lambda_1$ can exhibit a nonmonotonic behaviour as a function of the strength $\sigma$ of the off-diagonal matrix elements, as illustrated in Figs.~\ref{fig:reEntranceEffect} and \ref{fig:reEntranceEffect_with_outlier}.   As a consequence,  antagonistic interactions that are neither too strong nor too weak can stabilise  linear dynamical systems when the diagonal entries $D_i$ are heterogeneous, see Fig.~\ref{fig:phaseDiagCorrelation}.  

The results in Figs.~\ref{fig:reEntranceEffect} and \ref{fig:reEntranceEffect_with_outlier} can also be understood perturbatively.    Indeed,  a perturbative expansion of the eigenvalues of   $\mathbf{A}$ in the parameter $\sigma$  starting from the diagonal case with  $\sigma=0$ leads to the expression (see App.~\ref{App:C}) 
\begin{equation}
\lambda_j(\sigma) = D_j + \sum^n_{i=1; (i\neq j)}\frac{J_{ij}J_{ji}}{ (D_j-D_i)}+ O(\sigma^3).
\end{equation}
For the leading eigenvalue  $j=1$ it holds that $D_1-D_i\geq 0$ for all values  $i\in [n]$, and hence the second term is negative whenever $J_{ij}J_{ji}<0$, leading to the initial decrease of the leading eigenvalue $\lambda_1$ in Fig.~\ref{fig:reEntranceEffect}.   For larger values of $\sigma$ we need to consider the higher order terms in the perturbative  expansion of $\lambda_1$, which  are in general positive leading to the nonmonotonic behaviour of $\lambda_1$ in Fig.~\ref{fig:reEntranceEffect}.

We discuss various interesting questions for future research, both from the mathematical and ecological point of view.

For distributions $p_D$ that are uniform and bimodal, we have obtained the analytical expressions Eqs.~(\ref{outlier_bimodal}) and (\ref{outlier_uniform}) for the eigenvalue outlier, which led us to conjecture   Eq.~(\ref{eq:general}) for general distributions $p_D$.   However, we have no proof  of this general expression for the eigenvalue outlier, and hence it will be interesting to construct a proof of this simple, generic formula in  future work.  

For the case of a uniform distribution $p_D$ of the diagonal elements we have obtained an analytical expression for $\lambda_1$, which is given either  by Eq.~(\ref{eq:lambda1Analyt2}) or Eq.~(\ref{outlier_uniform}), depending on the value of $\mu$, and in the case of $\tau=0$ we have  obtained a closed form expression for the boundary of the support of the spectral density $\rho$, which is given by Eq.~(\ref{eq:devCircular}) and which also holds for uniform $p_D$.  The peculiar  solvability of the  uniform disorder case is consistent with  results obtained recently in  Ref.~\cite{mergny2021stability}  for symmetric matrices  ($\tau=1$).    Reference \cite{mergny2021stability} shows that  when $D_j = a+bj/n$, with $a$ and $b$ arbitrary constants, and  when the entries $J_{ij}$ are complex-valued  and Gaussian distributed, then  an explicit expression for the   joint distribution of eigenvalues can be obtained.   Based on the results in the present paper, one may speculate   that these results are extendable  to the case of $\tau=0$, which will be interesting to explore in future work.

An important extension of the present work are models of the form
\begin{equation}
A_{ij} = (J_{ij}+\mu)C_{ij} + \delta_{i,j}D_i   \label{sparse}
\end{equation} 
where $C_{ij}$ is now the adjacency matrix of a random graph.   The case of random directed graphs with a prescribed degree distribution $p_{K^{\rm in},K^{\rm out}}$ of  indegrees and  outdegrees has been solved in Refs.~\cite{neri2016eigenvalue, neri2020linear, tarnowski2020universal}.  This is the sparse equivalent of the "$\tau=0$"-case,  and in fact the oriented and locally tree-like structure of random directed graphs   leads to   a decoupling similar to those of Eqs.~(\ref{integralResolventEquation}) and (\ref{eq:SupportEq}) in the "$\tau=0$"-case.   For this reason, random directed graphs are  analytically tractable, and  Refs.~\cite{neri2016eigenvalue, neri2020linear, tarnowski2020universal}  derived for the boundary of the spectrum an equation similar to  Eq.~(\ref{eq:OrD}), but with a prefactor  that is given by $(\sigma^2+\mu^2) \frac{\langle K^{\rm in}K^{\rm out}\rangle}{c}$, i.e., 
 \begin{equation}
    1 =  (\sigma^2 +\mu^2)\frac{\langle K^{\rm in}K^{\rm out}\rangle}{c} \int_{\mathbb{R}} {\rm d}x \; p_D(x) \frac{1}{|\lambda - x|^2}, \label{eq:OrD2}
 \end{equation}  
 and analogously, Refs.~\cite{neri2016eigenvalue, neri2020linear, tarnowski2020universal}  for the eigenvalue outliers an equation similar to (Eq.~\ref{eq:outlierDir}), viz., 
\begin{equation}
1 = \mu\frac{\langle K^{\rm in}K^{\rm out}\rangle}{c}  \int_{\mathbb{R}} {\rm d}x \; p_D(x)  \frac{1}{\lambda_{\rm isol} - x} . \label{eq:outlierDir}
\end{equation}
The case of antagonistic interactions ($\tau<0$) is considerably more challenging  as one needs to know the distribution of the  diagonal entries $[\mathbf{G}]_{ii}$  of the resolvent, which is non trivial in the sparse case, see Ref.~\cite{mambuca2020dynamical}.    Nevertheless, Ref.~\cite{mambuca2020dynamical} analysed the antagonistic case without diagonal disorder and found  that  systems with antagonistic interactions are significantly  more stable than systems with mutualistic and competitive interactions (in fact, in the limit $n\rightarrow \infty$ they are infinitely more stable).    Ref.~\cite{mambuca2020dynamical} did however not study the effect of diagonal disorder on system stability.

In an ecological setting, the Jacobian matrix of a set of randomly coupled Lotka-Volterra equations have a specific structure, viz., all elements in a row are multiplied by the population abundance so that 
\begin{equation}
A_{ij} = D_iJ_{ij}.   
\end{equation} 
The spectra of such random matrix ensembles have been studied in Refs.~\cite{stone2018feasibility, PopAbundance}, and it would  be interesting to study the stabilising effect of antagonistic interactions in this setup.

The question of stability is also relevant for the study of experimental systems, see e.g.~Refs.\cite{de1995energetics, neutel2002stability, jacquet2016no}, and the matrices studied in the present paper are null models for real-world world systems.        However, as  discussed in detail in  Ref.~\cite{moore2012energetic},  most ecological data on foodwebs is qualitative, and  obtaining quantative data in particular on the Jacobian, is challenging.

Other interesting applications of the theory developed in the present paper are the study of Turing patterns that are governed by randomly coupled chemical reactions, which in Fourier space involves  a random matrix with diagonal disorder~\cite{baron2020dispersal}.

Let us end the paper with a word of caution when using the present results to understand the dynamics of nonlinear systems. In a linear system there exist  one fixed point, i.e., $\vec{x}=0$, and the system parameters will not affect the existence and uniqueness of this fixed point. However, in nonlinear systems this is in general not the case, see e.g.~\cite{fyodorov2016nonlinear, PhysRevE.103.022201}, and therefore system stability can also be affected by  bifurcations that eliminate  fixed points.  Moreover, for certain problems, such as stability in ecosystems, the fixed point has to be feasible, which for ecosystems implies that all entries of the fixed point are nonnegative \cite{may2019stability}, and this also constitutes an interesting random matrix theory problem \cite{bizeul2021positive}.    Another issue is that   $\mathbf{A}$ is the Hessian matrix, which is in general different from the interaction matrix.  Nevertheless,  studies  in, among others,  ecology \cite{biroli2018marginally, roy2019numerical, 10.21468/SciPostPhys.12.1.013}  and neuroscience  \cite{chaos, kadmon2015transition}, show that nonlinear systems do exhibit regimes with one unique stationary fixed point and random matrix theory can provide insights on system stability in this  regime.  In the ecological context for symmetric interactions $J_{ij}=J_{ji}$, Refs.~\cite{biroli2018marginally} shows that May's stability argument, albeit in the symmetric setting,  applies when the number of extinct species is correctly taking into account, and the corresponding spectrum of the Jacobian is described by random matrix theory.      Moreover, note that for symmetric interactions  replica theory can be used to determine the leading eigenvalue of the Hessian, see Refs.~\cite{biroli2018marginally, 10.21468/SciPostPhys.12.1.013, 10.21468/SciPostPhys.12.1.016}, while for nonsymmetric interactions this is not possible.

When preparing the manuscript, we  became aware of the preprint \cite{fyodorov2021counting} that also studies the spectral properties of matrices of the type  defined in Sec.~\ref{sec:2}.    However, the paper \cite{fyodorov2021counting} discusses the case of $\tau>0$ for which interactions further destabilise fixed points, whereas we were interested in the potentially stabilising effect of interactions for $\tau<0$.   

\section*{Acknowledgements}
IN thanks Jean-Philippe Bouchaud for several useful email communications, and  Andrea Mambuca,  Chiara Cammarota, and Pietro Valigi for fruitful discussions.     SC thanks Joseph Baron for pointing out Reference \cite{barabas2017self}, and  SC thanks Giorgio Carugno for a fruitful discussion.
  
\begin{appendix}

\section{Derivation of the generalised resolvent
equation~(\ref{integralResolventEquation}) using the Schur complement formula}\label{AppA}
We derive the Eq.~(\ref{integralResolventEquation}) using two useful properties.     

First, we use that permutation of a matrix and matrix inversion are two commutable operations.  Indeed, let $\mathbf{P}$ be the orthogonal matrix that represents an arbitrary permutation of the integers $\left\{1,2,\ldots,n\right\}$, then  
\begin{equation}
   \mathbf{P} \mathbf{H}^{-1} \mathbf{P}^{-1} = \left(  \mathbf{P} \mathbf{H} \mathbf{P}^{-1}\right)^{-1}. \label{eq:comm}
\end{equation}  
We use this property to perform the permutation \cite{metz2019spectral, tarnowski2020universal}
\begin{eqnarray}
[\mathbf{H}]_{j,k} \rightarrow  \left\{ \begin{array}{ccc}\left[\mathbf{\tilde{H}}\right]_{2j-1,2k-1}&{\rm  if}& 1\leq j,k\leq n, \\  \left[\mathbf{\tilde{H}}\right]_{2j-n,2k-1} &{\rm  if}& n+1\leq j \leq  2n, \: 1\leq k \leq  n, \\ \left[\mathbf{\tilde{H}}\right]_{2j-1,2k-n} &{\rm  if}&1\leq j \leq  n, \: n+1\leq k \leq  2n, \\  \left[\mathbf{\tilde{H}}\right]_{2j-n,2k-n} &{\rm  if}& n+1\leq j,k \leq  2n,  \end{array} \right.
\end{eqnarray}  
where $\tilde{\mathbf{H}}$ is the  matrix of permuted entries of $\mathbf{H}$.     The effect of this permutation is to bundle together the elements of $\mathbf{H}$ that depend on pairs of entries $(A_{ij},A_{ji})$.   

Second, we use the Schur inversion formula for the inverse of a  $2\times 2$ block matrix \cite{bun2017cleaning, metz2019spectral}, 
\begin{equation}
\left(\begin{array}{cc} \mathbf{a} & \mathbf{b} \\ \mathbf{c} & \mathbf{d} \end{array}\right) = \left(\begin{array}{cc} \mathbf{s}^{-1}_{\mathbf{d}} & -\mathbf{s}_{\mathbf{d}} \mathbf{b} \mathbf{d}^{-1} \\ -\mathbf{d}^{-1} \mathbf{c} \mathbf{s}_{\mathbf{d}} &  \mathbf{s}^{-1}_{\mathbf{a}} \end{array}\right),  \label{eq:schur}
\end{equation}   
where $\mathbf{s}_{\mathbf{d}} = \mathbf{a} - \mathbf{b}\mathbf{d}^{-1}\mathbf{c}$  and $\mathbf{s}_{\mathbf{a}} = \mathbf{d} - \mathbf{c}\mathbf{a}^{-1}\mathbf{b}$ are  the  Schur complements of the blocks $\mathbf{d}$ and $\mathbf{a}$, respectively.   If we choose for $\mathbf{a}$ the upper diagonal $2\times 2$ block of $\mathbf{\tilde{H}}$, then we obtain 
\begin{equation}
\mathsf{G}_{11} = \left( \left(\begin{array}{cc} \eta &  z-D_1\\  z^\ast-D_1&  \eta \end{array}\right) - \frac{1}{n}\sum^n_{j=2} \sum^n_{\ell=2}\left(\begin{array}{cc} 0 &  J_{1j}\\  J_{j1}&  0 \end{array}\right) \mathsf{G}^{(1)}_{j\ell}\left(\begin{array}{cc} 0 & J_{\ell 1}\\  J_{1\ell}&  0 \end{array}\right) \right)^{-1} ,
\end{equation}
where $\mathsf{G}^{(1)}_{j\ell}$ is defined as in Eq.~(\ref{eq:14}), but for a matrix $\mathbf{A}^{(1)}$ obtained by deleting the $1$-th row and $1$-th column of $\mathbf{A}$.
Permuting the entries of the matrix so that the $1$-th row and $1$-th  column are swapped with the $i$-th row and $i$-th  column, and using again Eq.~(\ref{eq:comm}), we obtain the analogous formula 
\begin{equation}
\mathsf{G}_{ii} = \left( \left(\begin{array}{cc} \eta &  z-D_i\\  z^\ast-D_i&  \eta \end{array}\right) - \frac{1}{n}\sum^n_{j=1;j\neq i}\:  \sum^n_{\ell=1;\ell\neq i} \left(\begin{array}{cc} 0 &  J_{ij}\\  J_{ji}&  0 \end{array}\right) \mathsf{G}^{(i)}_{j\ell}\left(\begin{array}{cc} 0 & J_{\ell i}\\  J_{i\ell}&  0 \end{array}\right) \right)^{-1} . \label{eq:G}
\end{equation}

Note that the sum in Eq.~(\ref{eq:G}) that runs over the indices $\ell$ and $j$ contains a very large number of terms in the limit of $n\gg1$.   Assuming that the law of large numbers applies to this sum ---  which can be verified to be the case, see Sec.~3 of Ref.~\cite{bai2008methodologies} --- we  replace the sum by its average value leading to
\begin{eqnarray}
\lefteqn{\sum^n_{j=1;j\neq i}\:  \sum^n_{\ell=1;\ell\neq i} \left(\begin{array}{cc} 0 &  J_{ij}\\  J_{ji}&  0 \end{array}\right) \mathsf{G}^{(i)}_{j\ell}\left(\begin{array}{cc} 0 & J_{\ell i}\\  J_{i\ell}&  0 \end{array}\right)}&&  \nonumber\\ 
&=& (n-1) \left(\begin{array}{cc} \langle J^2_{iu}\rangle \langle [\mathsf{G}^{(i)}_{uu}]_{22} \rangle  & \langle J_{iu}J_{ui}\rangle \langle [\mathsf{G}^{(i)}_{uu}]_{21}\rangle \\ \langle J_{iu}J_{ui}\rangle \langle [\mathsf{G}^{(i)}_{uu}]_{12}\rangle& \langle J^2_{ui}\rangle \langle [\mathsf{G}^{(i)}_{uu}]_{11}\rangle\end{array}\right)  
\nonumber\\ 
&& 
+ (n-1)(n-2) \left(\begin{array}{cc} \langle J_{iu}\rangle \langle J_{iv }\rangle \langle [\mathsf{G}^{(i)}_{uv}]_{22} \rangle  & \langle J_{iu}\rangle \langle J_{v i}\rangle \langle [\mathsf{G}^{(i)}_{uv}]_{21}\rangle \\ \langle J_{iv}\rangle \langle J_{ui }\rangle  \langle [\mathsf{G}^{(i)}_{uv}]_{12}\rangle& \langle J_{ui}\rangle \langle J_{v i}\rangle \langle [\mathsf{G}^{(i)}_{uv}]_{11}\rangle\end{array}\right)  ,  \label{eq:large}
\end{eqnarray} 
with $u,v\in [n]\setminus \left\{i\right\}$ and $u\neq v$.  In Eq.~(\ref{eq:large}) we have used that  the pairs $(J_{ij},J_{ji})$ are identically and independently distributed random variables.    Moreover, we have  used that $\langle [\mathsf{G}^{(i)}_{uu}]_{12}\rangle$  and $\langle [\mathsf{G}^{(i)}_{uv}]_{12}\rangle$  are independent of $i$, $u$ and $v$, as in the definition of the random matrix  model for $\mathbf{A}$ all indices are equivalent.   Since $\langle J_{iu}\rangle=0$,  the second term in Eq.~(\ref{eq:large}) equals zero, and using that $\sigma^2 = \langle J^2_{ik}\rangle$ and $\tau \sigma^2 = \langle J_{ik}J_{ki}\rangle$, we obtain 
\begin{equation}
 \mathsf{G}_{ii} =  \left( \left(\begin{array}{cc} \eta &  z-D_i\\  z^\ast-D_i&  \eta \end{array}\right) - \sigma^2 \left(\begin{array}{cc}  \langle [\mathsf{G}^{(i)}_{uu}]_{22} \rangle  & \tau \langle [\mathsf{G}^{(i)}_{uu}]_{21}\rangle \\ \tau  \langle [\mathsf{G}^{(i)}_{uu}]_{12}\rangle&  \langle [\mathsf{G}^{(i)}_{uu}]_{11}\rangle\end{array}\right)   \right)^{-1}+ o_n(1), \label{eq:GiiApp}
\end{equation}
where $o(\cdot)$ is  the small o notation. Taking the ensemble average of Eq.~(\ref{eq:GiiApp}) and, in the limit of large $n$, identifying 
\begin{equation}
\mathsf{g} = \langle  \mathsf{G}_{ii}  \rangle  
\end{equation}
for all $i\in[n]$, 
and 
\begin{equation}
\mathsf{g} = \langle  \mathsf{G}^{(i)}_{uu}   \rangle  \label{eq:gnM1} 
\end{equation}
for all $u\in[n]\setminus\left\{i\right\}$, we obtain    Eq.~(\ref{integralResolventEquation}).   Equation~(\ref{eq:gnM1}) follows from the fact that  $\mathbf{A}^{(i)}$ is drawn from the same ensemble as $\mathbf{A}$, except that $n\rightarrow n-1$.

\section{Derivation of Eq.~(\ref{eq:devCircular}) for the boundary of the spectrum when $\tau=0$ and $p_D$ is uniform}\label{app:B}
Equation~(\ref{eq:OrD}) for the uniform distribution Eq.~(\ref{eq:uniform}) gives 
\begin{equation}
\int_{d_-}^{d_+}\frac{1}{(x-u)^2 + y^2}\mathrm{d}u=\frac{d_+-d_-}{\sigma^2},
\end{equation}
where we have used  $z = x+{\rm i}y$.    Using the formula 
\begin{equation}
\int \frac{1}{x^{2}+a^{2}} d x=\frac{1}{a} \arctan \frac{x}{a}+{\rm constant}
\end{equation} 
for the indefinite integral of $1/(x^2+a^2)$ when $a\neq 0$, we obtain  that for $y\neq 0$
\begin{equation}
\arctan\left[\frac{(d_+ - x)}{y}\right]-\arctan\left[\frac{(d_- - x)}{y}\right]=\frac{(d_+-d_-)}{\sigma^2}y.\label{eq:atanhs}
\end{equation} 
Notice that since $\arctan(a)\in (-\pi/2,\pi/2)$, we have the condition $y\in  (-\pi\sigma^2/(d_+-d_-),\pi\sigma^2/(d_+-d_-)$.  Subsequently, using 
\begin{equation}
\arctan a-\arctan b=\arctan \left(\frac{a-b}{1+a b}\right)
\end{equation} 
 in Eq.~(\ref{eq:atanhs}), we obtain Eq.~(\ref{eq:devCircular}).
 
 \section{Derivation of Eq.~(\ref{eq:lambda1Analyt2}) for the leading eigenvalue $\lambda_1$ in the case of uniformly distributed diagonal elements}
\label{AppD}
We derive  Eq.~(\ref{eq:lambda1Analyt2}) for the leading eigenvalue $\lambda_1$ when $p_D$ is the uniform distribution given by Eq.~(\ref{eq:uniform}).

Using the assumption that the leading eigenvalue $\lambda_1\in \mathbb{R}$, we  set $y=0$ in equations Eqs.~(\ref{re_g12}-\ref{eq:boundary_final}) yielding
 \begin{equation}
   \Re(g_{12}) =-\left\langle\frac{1}{(D-x)+\Re(g_{12}) \tau \sigma^2}\right\rangle_D
    \label{re_g12_y_is_0}
\end{equation}
and
\begin{equation}
    1=\left\langle\frac{\sigma^2}{\left[(D-x)+\Re(g_{12}) \tau \sigma^2\right]^2}\right\rangle_D.
    \label{eq:boundary_final_y_is_0}
\end{equation} 

Integrating the equations  Eq.~(\ref{re_g12_y_is_0}-\ref{eq:boundary_final_y_is_0})  over the uniform distribution $p_D$ supported on the interval $[d_-,d_+]$, we obtain
\begin{equation}
 \Re(g_{12}) =\frac{\log \left(x-d_- - \Re(g_{12}) \sigma ^2 \tau \right)}{d_+-d_-}-\frac{\log \left(x-d_+-\Re(g_{12}) \sigma ^2 \tau \right)}{d_+-d_-}
 \label{eq:uni_ave_1}
\end{equation}
and
\begin{equation}
1=\frac{\sigma^2}{(d_+-x+\Re(g_{12})\tau\sigma ^2 )(d_- - x+\Re(g_{12})\tau\sigma ^2 )}.
\label{eq:uni_ave_2}
\end{equation}
We first solve Eq.~(\ref{eq:uni_ave_2}) towards $ \Re(g_{12})$  with solutions
\begin{equation}
 \Re(g_{12})=\frac{x}{\sigma^2\tau}-\frac{  \left(d_-+d_+\right)}{2 \sigma ^2 \tau }+s\frac{ \sqrt{ \left(d_+-d_-\right) ^2+4 \sigma
   ^2}}{2 \sigma ^2 \tau },
   \label{g12_sol}
\end{equation}
where $s=\pm1$. Replacing $ \Re(g_{12})$ in Eq.~(\ref{eq:uni_ave_1}) by this solution gives a linear equation in $x$.    The solutions of this linear equation provide the intersection points of the boundary of the support of the spectral distribution with the  real axis, viz., 
\begin{eqnarray}
x&=&\frac{1}{2} \left(-s\sqrt{\left(d_--d_+\right){}^2+4 \sigma
   ^2}+d_-+d_+\right)  \nonumber\\ 
  && +\tau \frac{\sigma ^2 }{d_+-d_-} \log
   \left(\frac{-s\sqrt{\left(d_--d_+\right){}^2+4 \sigma
   ^2}+d_+ -d_-}{-s\sqrt{\left(d_--d_+\right){}^2+4 \sigma
   ^2}-d_++d_-}\right).
\end{eqnarray} 
For $s=1$ we obtain the leading eigenvalue given by Eq.~\eqref{eq:lambda1Analyt2}.
 \section{Perturbation theory for the leading eigenvalue }\label{App:C}
 We use perturbation theory to understand the functional behaviour in Fig.~\ref{fig:reEntranceEffect} of $\Re(\lambda_1)$ as a function of $\sigma$.  
 
 Let $\mathbf{D}+\sigma\mathbf{J}$,  where $\mathbf{D}$ is a diagonal matrix,  $\sigma$  a small parameter, and $\mathbf{J}$ an arbitrary $\sigma$-independent matrix with zero-valued diagonal entries.    Let $\lambda^{(0)}_j$, $\vec{r}^{(0)}_j$, and $\vec{l}^{(0)}_j$ denote the eigenvalues, right eigenvectors, and left eigenvectors of $\mathbf{D}$, respectively.     
 
 Let $\lambda_j(\sigma)$ denote the eigenvalues of  $\mathbf{D}+\sigma\mathbf{J}$.   An expansion around $\sigma\approx0$ gives
 \begin{equation}
 \lambda_j(\sigma) = \lambda^{(0)}_j  + \lambda^{(1)}_j \sigma + \lambda^{(2)}_j \sigma^2 + O(\sigma^3)
 \end{equation}
 with $\lambda^{(0)}_j = D_j$.    Note that in this paper we use the convention that $\lambda^{(0)}_1\geq \lambda^{(0)}_2 \geq \ldots \lambda^{(0)}_n$ and thus $D_1\geq D_2\geq \ldots D_n$.
 
It then holds  \cite{wilkinson1988algebraic} 
\begin{equation}
\lambda^{(1)}_j = \frac{\vec{l}^{(0)}_j\cdot \mathbf{J}\vec{r}^{(0)}_j}{\vec{l}^{(0)}_j\cdot \vec{r}^{(0)}_j}
\end{equation}
and 
\begin{equation}
\lambda^{(2)}_j = \frac{1}{\vec{l}^{(0)}_j\cdot \vec{r}^{(0)}_j}\sum^n_{i=1; i\neq j}\frac{[\vec{l}^{(0)}_j\cdot\mathbf{J}\vec{r}^{(0)}_i][\vec{l}^{(0)}_i\cdot\mathbf{J}\vec{r }^{(0)}_j]}{(\vec{l}^{(0)}_i\cdot \vec{r}^{(0)}_i) (\lambda^{(0)}_j-\lambda^{(0)}_i)}.
\end{equation} 

Since $\mathbf{D}$ is a diagonal matrix,  we can set $\vec{l}^{(0)}_i\cdot \vec{r}^{(0)}_i=\delta_{i,j}$ and $\vec{l}^{(0)}_j\cdot\mathbf{J}\vec{r}^{(0)}_i = J_{ji} (1-\delta_{i,j})$.   In this case, it holds that
\begin{equation}
\lambda_j(\sigma) = D_j + \sum^n_{i=1; i\neq j }\frac{J_{ij}J_{ji}}{ (D_j-D_i)}\sigma^2 + O(\sigma^3). \label{eq:lambdaD}
\end{equation} 
For $\sigma=0$, it holds that $D_1 =  D_{\rm max}$ and thus the denonimator in Eq.~(\ref{eq:lambdaD}) is positive.  From this it follows that when $J_{ij}J_{ji}<0$, $\lambda_1$ initially decreases as a function of $\sigma$.    However, when $\sigma$ is larger, then the $O(\sigma^3)$ becomes  relevant, which provides the nonmonotonic behaviour in Fig.~\ref{fig:phaseDiagCorrelation}. 

\section{Boundary of the spectrum when $\mu\neq0$} \label{app:muDiffZero}
We show that   the  support set    $\mathcal{S}$ of $\rho(z)$ is, in the limit of large $n$, independent of $\mu$, and hence in this limit the boundary of $\mathcal{S}$  
solves the Eqs.~(\ref{integralResolventEquation}) and (\ref{eq:SupportEq}).   

Following the derivation of Appendix~\ref{AppA}, we obtain, instead of the self-consistent  Eq.~(\ref{eq:G}) for $\mathsf{G}_{ii}$, the  self-consistent equation  
\begin{eqnarray}
\mathsf{G}_{ii} &=& \left( \left(\begin{array}{cc} \eta &  z-D_i+\mu/n\\  z^\ast-D_i+\mu/n&  \eta \end{array}\right) - \right.  
\nonumber\\ 
&& \left.
\frac{1}{n}\sum^n_{j=1;j\neq i}\:  \sum^n_{\ell=1;\ell\neq i} \left(\begin{array}{cc} 0 &  J_{ij}+\mu/n\\  J_{ji}+\mu/n&  0 \end{array}\right) \mathsf{G}^{(i)}_{j\ell}\left(\begin{array}{cc} 0 & J_{\ell i}+\mu/n\\  J_{i\ell}+\mu/n&  0 \end{array}\right) \right)^{-1} .\nonumber\\ \label{eq:G2}
\end{eqnarray}  
Assuming the law of large numbers applies, we obtain for the second term in the previous equation, 
\begin{eqnarray}
\lefteqn{\sum^n_{j=1;j\neq i}\:  \sum^n_{\ell=1;\ell\neq i} \left(\begin{array}{cc} 0 &  J_{ij}+\mu/n\\  J_{ji}+\mu/n&  0 \end{array}\right) \mathsf{G}^{(i)}_{j\ell}\left(\begin{array}{cc} 0 & J_{\ell i}+\mu/n\\  J_{i\ell}+\mu/n&  0 \end{array}\right)}&&  \nonumber\\ 
&=& (n-1) \left(\begin{array}{cc} \left(\langle J^2_{iu}\rangle + \frac{\mu^2}{n^2}\right)\langle [\mathsf{G}^{(i)}_{uu}]_{22} \rangle  & \left(\langle J_{iu}J_{ui}\rangle + \frac{\mu^2}{n^2}\right) \langle [\mathsf{G}^{(i)}_{uu}]_{21}\rangle \\ \left(\langle J_{iu}J_{ui}\rangle+ \frac{\mu^2}{n^2}\right)  \langle [\mathsf{G}^{(i)}_{uu}]_{12}\rangle& \left(\langle J^2_{ui}\rangle + \frac{\mu^2}{n^2}\right) \langle [\mathsf{G}^{(i)}_{uu}]_{11}\rangle\end{array}\right)  
\nonumber\\ 
&& 
+ \frac{(n-1)(n-2)}{n^2}\mu^2 \left(\begin{array}{cc} \langle  [\mathsf{G}^{(i)}_{uv}]_{22} \rangle  &  \langle [\mathsf{G}^{(i)}_{uv}]_{21}\rangle \\   \langle [\mathsf{G}^{(i)}_{uv}]_{12}\rangle&  \langle [\mathsf{G}^{(i)}_{uv}]_{11}\rangle\end{array}\right)   \label{eq:large2}
\end{eqnarray} 
with $u,v\in [n]\setminus \left\{i\right\}$ and $u\neq v$.  
Using that $\sigma^2 = \langle J^2_{ik}\rangle$ and $\tau \sigma^2 = \langle J_{ik}J_{ki}\rangle$, Eqs.~(\ref{eq:G2}) and (\ref{eq:large2}) yield Eq.~(\ref{eq:GiiApp}), and consequently, in the limit of large $n$ the quantity $\mathsf{G}_{ii}$ that determines the resolvent of $\mathbf{A}$ is, neglecting subleading order terms in $n$, independent of~$\mu$.

\section{Derivation of the
Eq.~(\ref{eq_outlierx}) for the eigenvalue outlier $\lambda_{\rm isol}$}\label{app:outlier2}  
We derive Eq.~(\ref{eq_outlierx})  for the eigenvalue outlier of random matrices $\mathbf{A}$ as defined in Sec.~\ref{sec:2}.    To this purpose, we use the cavity method for eigenvalue outliers of random matrices, as   developed in Refs.~\cite{neri2016eigenvalue, neri2020linear, metz2021localization}, and  used in Ref.~\cite{giorgio} for the case of symmetric block matrices.   Note that this method is distinct from the cavity method for the spectral density developed in \cite{rogers2009cavity}.

Assuming $\mathbf{A}$ is diagonalisable, the matrix $\mathbf{A}$ has $n$ left and right eigenvectors denoted, respectively, by   $\vec{l}_j$ and $\vec{r}_j$.   Normalising left and right eigenvectors such that 
\begin{equation}
    \vec{l}^\dagger_j \: \vec{r}_k = \delta_{j,k} , \quad \forall j,k\in[n],
\end{equation}
we can decompose the matrix as
\begin{equation}
\mathbf{A} =    \sum^n_{j=1}  \lambda_j \vec{r}_j\:  \vec{l}^\dagger_j,
\end{equation}
and analogously for the resolvent
\begin{equation}
\mathbf{G}(z) = \sum^n_{j=1}  \frac{1}{z-\lambda_j} \vec{r}_j\:  \vec{l}^\dagger_j, \quad {\rm for} \quad z\notin \left\{\lambda_1, \lambda_2,\ldots, \lambda_n\right\}. 
\end{equation}
Consequently, for $z = \lambda_j + \eta$ it holds that 
\begin{equation}
\lim_{\eta\rightarrow 0}    \eta  \mathbf{G}(\lambda_j+\eta) =   \vec{r}_j\:  \vec{l}^\dagger_j + O(\eta),    \label{eq:outlierFx}
\end{equation} 
as long as $\eta$ is much smaller than the distance between $\lambda_j$ and any other eigenvalue of $\mathbf{A}$.    
Using the Schur complement formula Eq.~(\ref{eq:schur}), the  Appendix F of Ref.~\cite{neri2020linear} shows that the eigenvector entries $[\vec{r}_j]_k $ of $\vec{r}_j$ obey the recursion relation
\begin{equation}
[\vec{r}_j]_k =  G_{kk}(\lambda_j+\eta) \sum^n_{\ell=1;(\ell\neq k)}A_{k\ell}[\vec{r}^{(k)}_j]_{\ell} \label{eq:Gk}
\end{equation}   
in the asymptotic limit $n\gg 1$,
where $\vec{r}^{(k)}_j$ is the right eigenvector of the matrix $\mathbf{A}^{(k)}$, obtained from $\mathbf{A}$ by deleting the $k$-th rows and columns, and associated with the same eigenvalue $\lambda_j$.    
The derivation of Eq.~(\ref{eq:Gk}) relies on two assumptions, mainly that the eigenvalue $\lambda_j$ is well separated from other eigenvalues in the limit of large $n$, so that Eq.~(\ref{eq:outlierFx}) applies, and that $\lambda_j$ is both an eigenvalue of   $\mathbf{A}$ and the cavity matrices $\mathbf{A}^{(k)}$.   Both assumptions  are valid for eigenvalue outliers $\lambda_{\rm isol}$ in the limit of large $n$.

Note that the $[\vec{r}_j]_k$, just as the $G_{kk}(\lambda_j+\eta) $, are fluctuating quantities.      However, by setting $\lambda_j=\lambda_{\rm isol}$ and taking the ensemble average on the right and left hand side of Eq.~(\ref{eq:Gk2}) we obtain an equation for the eigenvalue outliers $\lambda_{\rm isol}$.   
Indeed, using the law of large numbers,
\begin{equation}
\lim_{n\rightarrow \infty}\sum^n_{\ell=1;(\ell\neq j)}A_{k\ell}[\vec{r}^{(k)}_j]_{\ell} = \mu \langle R_{\rm isol}\rangle,  \label{eq:86}
\end{equation}   
where $\langle R_{\rm isol}\rangle\neq 0$ is the average value of the entries of the right eigenvector associated with the eigenvalue outlier, i.e., $\langle [\vec{r}_j]_k\rangle =  \langle [\vec{r}^{(k)}_j]_{\ell}\rangle = \langle R_{\rm isol}\rangle$.     Substitution in (\ref{eq:Gk}) and taking the ensemble average yields 
\begin{equation}
\langle R_{\rm isol}\rangle =  \mu \langle R_{\rm isol}\rangle g_{21}(\lambda_{\rm isol}),  \label{eq:Gk2}
\end{equation}   
where $ g_{21}(\lambda_{\rm isol}) = \lim_{\eta\rightarrow 0}\langle G_{kk}(\lambda_{\rm isol}+\eta)\rangle$ solves the Eq.~(\ref{integralResolventEquation}) for $g_{22} = g_{11} = 0$ as the outlier is in the region of the complex plane
outside the support set $\mathcal{S}$ and hence the trivial solution applies.
As for eigenvalue outliers $\langle R_{\rm isol}\rangle \neq 0$, Eq.~(\ref{eq:Gk2})  implies that  the outlier eigenvalue $\lambda_{\rm isol}$ solves  the  Eq.~(\ref{eq_outlierx}).

\end{appendix}

\bibliography{SciPost_LaTeX_Template}

\begin{thebibliography}{10}
\providecommand{\url}[1]{\texttt{#1}}
\providecommand{\urlprefix}{URL }
\expandafter\ifx\csname urlstyle\endcsname\relax
  \providecommand{\doi}[1]{doi:\discretionary{}{}{}#1}\else
  \providecommand{\doi}{doi:\discretionary{}{}{}\begingroup
  \urlstyle{rm}\Url}\fi
\providecommand{\eprint}[2][]{\url{#2}}

\bibitem{may2019stability}
R.~M. May,
\newblock \emph{Stability and complexity in model ecosystems},
\newblock In \emph{Stability and Complexity in Model Ecosystems}. Princeton
  university press,
\newblock \doi{10.1515/9780691206912} (2001).

\bibitem{chaos}
H.~Sompolinsky, A.~Crisanti and H.~J. Sommers,
\newblock \emph{Chaos in random neural networks},
\newblock Physical Review Letters \textbf{61}, 259 (1988),
\newblock \doi{10.1103/PhysRevLett.61.259}.

\bibitem{kadmon2015transition}
J.~Kadmon and H.~Sompolinsky,
\newblock \emph{Transition to chaos in random neuronal networks},
\newblock Phys. Rev. X \textbf{5}, 041030 (2015),
\newblock \doi{10.1103/PhysRevX.5.041030}.

\bibitem{ahmadian2015properties}
Y.~Ahmadian, F.~Fumarola and K.~D. Miller,
\newblock \emph{Properties of networks with partially structured and partially
  random connectivity},
\newblock Physical Review E \textbf{91}(1), 012820 (2015),
\newblock \doi{10.1103/PhysRevE.91.012820}.

\bibitem{quirk1965qualitative}
J.~Quirk and R.~Ruppert,
\newblock \emph{Qualitative economics and the stability of equilibrium},
\newblock The review of economic studies \textbf{32}(4), 311 (1965),
\newblock \doi{10.2307/2295838}.

\bibitem{wigner1958distribution}
E.~P. Wigner,
\newblock \emph{On the distribution of the roots of certain symmetric
  matrices},
\newblock Annals of Mathematics pp. 325--327 (1958),
\newblock \doi{10.2307/1970008}.

\bibitem{dyson1962statistical}
F.~J. Dyson,
\newblock \emph{Statistical theory of the energy levels of complex systems. i},
\newblock Journal of Mathematical Physics \textbf{3}(1), 140 (1962),
\newblock \doi{10.1063/1.1703773}.

\bibitem{may1972will}
R.~M. May,
\newblock \emph{Will a large complex system be stable?},
\newblock Nature \textbf{238}(5364), 413 (1972),
\newblock \doi{10.1038/238413a0}.

\bibitem{allesina2012stability}
S.~Allesina and S.~Tang,
\newblock \emph{Stability criteria for complex ecosystems},
\newblock Nature \textbf{483}(7388), 205 (2012),
\newblock \doi{10.1038/nature10832}.

\bibitem{mougi2012diversity}
A.~Mougi and M.~Kondoh,
\newblock \emph{Diversity of interaction types and ecological community
  stability},
\newblock Science \textbf{337}(6092), 349 (2012),
\newblock \doi{10.1126/science.1220529}.

\bibitem{mougi2014stability}
A.~Mougi and M.~Kondoh,
\newblock \emph{Stability of competition--antagonism--mutualism hybrid
  community and the role of community network structure},
\newblock Journal of theoretical biology \textbf{360}, 54 (2014),
\newblock \doi{10.1016/j.jtbi.2014.06.030}.

\bibitem{biroli2018marginally}
G.~Biroli, G.~Bunin and C.~Cammarota,
\newblock \emph{Marginally stable equilibria in critical ecosystems},
\newblock New Journal of Physics \textbf{20}(8), 083051 (2018),
\newblock \doi{10.1088/1367-2630/aada58}.

\bibitem{roy2019numerical}
F.~Roy, G.~Biroli, G.~Bunin and C.~Cammarota,
\newblock \emph{Numerical implementation of dynamical mean field theory for
  disordered systems: Application to the lotka--volterra model of ecosystems},
\newblock Journal of Physics A: Mathematical and Theoretical \textbf{52}(48),
  484001 (2019),
\newblock \doi{10.1088/1751-8121/ab1f32}.

\bibitem{10.21468/SciPostPhys.12.1.013}
A.~Altieri and G.~Biroli,
\newblock \emph{{Effects of intraspecific cooperative interactions in large
  ecosystems}},
\newblock SciPost Phys. \textbf{12}, 013 (2022),
\newblock \doi{10.21468/SciPostPhys.12.1.013}.

\bibitem{moran2019may}
J.~Moran and J.-P. Bouchaud,
\newblock \emph{May's instability in large economies},
\newblock Phys. Rev. E \textbf{100}, 032307 (2019),
\newblock \doi{10.1103/PhysRevE.100.032307}.

\bibitem{guo2021exploring}
Y.~Guo and A.~Amir,
\newblock \emph{Exploring the effect of network topology, mrna and protein
  dynamics on gene regulatory network stability},
\newblock Nature communications \textbf{12}(1), 1 (2021),
\newblock \doi{10.1038/s41467-020-20472-x}.

\bibitem{neri2020linear}
I.~Neri and F.~L. Metz,
\newblock \emph{Linear stability analysis of large dynamical systems on random
  directed graphs},
\newblock Phys. Rev. Research \textbf{2}, 033313 (2020),
\newblock \doi{10.1103/PhysRevResearch.2.033313}.

\bibitem{tarnowski2020universal}
W.~Tarnowski, I.~Neri and P.~Vivo,
\newblock \emph{Universal transient behavior in large dynamical systems on
  networks},
\newblock Phys. Rev. Research \textbf{2}, 023333 (2020),
\newblock \doi{10.1103/PhysRevResearch.2.023333}.

\bibitem{mambuca2020dynamical}
A.~M. Mambuca, C.~Cammarota and I.~Neri,
\newblock \emph{Dynamical systems on large networks with predator-prey
  interactions are stable and exhibit oscillations},
\newblock Phys. Rev. E \textbf{105}, 014305 (2022),
\newblock \doi{10.1103/PhysRevE.105.014305}.

\bibitem{cicuta2016random}
G.~M. Cicuta and L.~G. Molinari,
\newblock \emph{Random antagonistic matrices},
\newblock Journal of Physics A: Mathematical and Theoretical \textbf{49}(37),
  375601 (2016),
\newblock \doi{10.1088/1751-8113/49/37/375601}.

\bibitem{girko1986elliptic}
V.~Girko,
\newblock \emph{Elliptic law},
\newblock Theory of Probability \& Its Applications \textbf{30}(4), 677 (1986),
\newblock \doi{10.1137/1130089}.

\bibitem{sommers1988spectrum}
H.~J. Sommers, A.~Crisanti, H.~Sompolinsky and Y.~Stein,
\newblock \emph{Spectrum of large random asymmetric matrices},
\newblock Phys. Rev. Lett. \textbf{60}, 1895 (1988),
\newblock \doi{10.1103/PhysRevLett.60.1895}.

\bibitem{PhysRevLett.60.1895}
H.~J. Sommers, A.~Crisanti, H.~Sompolinsky and Y.~Stein,
\newblock \emph{Spectrum of large random asymmetric matrices},
\newblock Phys. Rev. Lett. \textbf{60}, 1895 (1988),
\newblock \doi{10.1103/PhysRevLett.60.1895}.

\bibitem{Pastur}
L.~A. Pastur,
\newblock \emph{On the spectrum of random matrices},
\newblock Theoretical and Mathematical Physics \textbf{10}(1), 67 (1972),
\newblock \doi{10.1007/BF01035768}.

\bibitem{lee2016extremal}
J.~O. Lee and K.~Schnelli,
\newblock \emph{Extremal eigenvalues and eigenvectors of deformed wigner
  matrices},
\newblock Probability Theory and Related Fields \textbf{164}(1-2), 165 (2016),
\newblock \doi{10.1007/s00440-014-0610-8}.

\bibitem{mergny2021stability}
P.~Mergny and S.~N. Majumdar,
\newblock \emph{Stability of large complex systems with heterogeneous
  relaxation dynamics},
\newblock Journal of Statistical Mechanics: Theory and Experiment
  \textbf{2021}(12), 123301 (2021),
\newblock \doi{10.1088/1742-5468/ac3b47}.

\bibitem{neri2016eigenvalue}
I.~Neri and F.~L. Metz,
\newblock \emph{Eigenvalue outliers of non-hermitian random matrices with a
  local tree structure},
\newblock Phys. Rev. Lett. \textbf{117}, 224101 (2016),
\newblock \doi{10.1103/PhysRevLett.117.224101}.

\bibitem{khoruzhenko1996large}
B.~Khoruzhenko,
\newblock \emph{Large-n eigenvalue distribution of randomly perturbed
  asymmetric matrices},
\newblock Journal of Physics A: Mathematical and General \textbf{29}(7), L165
  (1996),
\newblock \doi{10.1088/0305-4470/29/7/003}.

\bibitem{barabas2017self}
G.~Barab{\'a}s, M.~J. Michalska-Smith and S.~Allesina,
\newblock \emph{Self-regulation and the stability of large ecological
  networks},
\newblock Nature ecology \& evolution \textbf{1}(12), 1870 (2017),
\newblock \doi{10.1038/s41559-017-0357-6}.

\bibitem{bai2008methodologies}
Z.~D. Bai,
\newblock \emph{Methodologies in spectral analysis of large dimensional random
  matrices, a review},
\newblock In \emph{Advances in statistics}, pp. 174--240. World Scientific,
\newblock \doi{10.1142/9789812793096_0015} (2008).

\bibitem{tao2012topics}
T.~Tao,
\newblock \emph{Topics in random matrix theory}, vol. 132,
\newblock American Mathematical Soc.,
\newblock \doi{10.1090/gsm/132} (2012).

\bibitem{tao2013outliers}
T.~Tao,
\newblock \emph{Outliers in the spectrum of iid matrices with bounded rank
  perturbations},
\newblock Probability Theory and Related Fields \textbf{155}(1), 231 (2013),
\newblock \doi{10.1007/s00440-011-0397-9}.

\bibitem{o2014low}
S.~O'Rourke and D.~Renfrew,
\newblock \emph{Low rank perturbations of large elliptic random matrices},
\newblock Electronic Journal of Probability \textbf{19}, 1 (2014),
\newblock \doi{10.1214/EJP.v19-3057}.

\bibitem{feinberg1997non}
J.~Feinberg and A.~Zee,
\newblock \emph{Non-hermitian random matrix theory: Method of hermitian
  reduction},
\newblock Nuclear Physics B \textbf{504}(3), 579 (1997),
\newblock \doi{10.1016/S0550-3213(97)00502-6}.

\bibitem{hikami1998density}
S.~Hikami and R.~Pnini,
\newblock \emph{Density of state in a complex random matrix theory with
  external source},
\newblock Journal of Physics A: Mathematical and General \textbf{31}(35), L587
  (1998),
\newblock \doi{10.1088/0305-4470/31/35/001}.

\bibitem{rogers2009cavity}
T.~Rogers and I.~P. Castillo,
\newblock \emph{Cavity approach to the spectral density of non-hermitian sparse
  matrices},
\newblock Physical Review E \textbf{79}(1), 012101 (2009),
\newblock \doi{10.1103/physreve.79.012101}.

\bibitem{metz2019spectral}
F.~L. Metz, I.~Neri and T.~Rogers,
\newblock \emph{Spectral theory of sparse non-hermitian random matrices},
\newblock Journal of Physics A: Mathematical and Theoretical  (2019),
\newblock \doi{10.1088/1751-8121/ab1ce0}.

\bibitem{bai2010spectral}
Z.~Bai and J.~W. Silverstein,
\newblock \emph{Spectral analysis of large dimensional random matrices},
  vol.~20,
\newblock Springer,
\newblock \doi{10.1007/978-1-4419-0661-8} (2010).

\bibitem{bordenave2012around}
C.~Bordenave and D.~Chafa{\"\i},
\newblock \emph{Around the circular law},
\newblock Probability surveys \textbf{9}, 1 (2012),
\newblock \doi{10.1214/11-PS183}.

\bibitem{wood2012universality}
P.~M. Wood,
\newblock \emph{Universality and the circular law for sparse random matrices},
\newblock The Annals of Applied Probability \textbf{22}(3), 1266 (2012),
\newblock \doi{10.1214/11-aap789}.

\bibitem{cook2018non}
N.~Cook, W.~Hachem, J.~Najim and D.~Renfrew,
\newblock \emph{Non-hermitian random matrices with a variance profile (i):
  deterministic equivalents and limiting esds},
\newblock Electronic Journal of Probability \textbf{23}, 1 (2018),
\newblock \doi{10.1214/18-EJP230}.

\bibitem{cook2019circular}
N.~Cook,
\newblock \emph{The circular law for random regular digraphs},
\newblock In \emph{Annales de l'Institut Henri Poincar{\'e}, Probabilit{\'e}s
  et Statistiques}, vol.~55, pp. 2111--2167. Institut Henri Poincar{\'e},
\newblock \doi{10.1214/18-AIHP943} (2019).

\bibitem{livan2018introduction}
G.~Livan, M.~Novaes and P.~Vivo,
\newblock \emph{Introduction to random matrices: theory and practice}, vol.~26,
\newblock Springer,
\newblock \doi{10.1007/978-3-319-70885-0} (2018).

\bibitem{girko1985circular}
V.~L. Girko,
\newblock \emph{Circular law},
\newblock Theory of Probability \& Its Applications \textbf{29}(4), 694 (1985),
\newblock \doi{10.1137/1129095}.

\bibitem{leadbetter2012extremes}
{M R Leadbetter, Georg Lindgren, and Holger Rootz{\'e}n},
\newblock \emph{Extremes and related properties of random sequences and
  processes},
\newblock Springer-Verlag, 1st edn.,
\newblock \doi{10.1007/978-1-4612-5449-2} (1983).

\bibitem{gardner1970connectance}
M.~R. Gardner and W.~R. Ashby,
\newblock \emph{Connectance of large dynamic (cybernetic) systems: critical
  values for stability},
\newblock Nature \textbf{228}(5273), 784 (1970),
\newblock \doi{10.1038/228784a0}.

\bibitem{PhysRevResearch.2.043116}
F.~L. Metz and J.~D. Silva,
\newblock \emph{Spectral density of dense random networks and the breakdown of
  the wigner semicircle law},
\newblock Phys. Rev. Research \textbf{2}, 043116 (2020),
\newblock \doi{10.1103/PhysRevResearch.2.043116}.

\bibitem{stone2018feasibility}
L.~Stone,
\newblock \emph{The feasibility and stability of large complex biological
  networks: a random matrix approach},
\newblock Scientific reports \textbf{8}(1), 1 (2018),
\newblock \doi{10.1038/s41598-018-26486-2}.

\bibitem{PopAbundance}
T.~Gibbs, J.~Grilli, T.~Rogers and S.~Allesina,
\newblock \emph{Effect of population abundances on the stability of large
  random ecosystems},
\newblock Phys. Rev. E \textbf{98}, 022410 (2018),
\newblock \doi{10.1103/PhysRevE.98.022410}.

\bibitem{de1995energetics}
P.~C. De~Ruiter, A.-M. Neutel and J.~C. Moore,
\newblock \emph{Energetics, patterns of interaction strengths, and stability in
  real ecosystems},
\newblock Science \textbf{269}(5228), 1257 (1995),
\newblock \doi{10.1126/science.269.5228.1257}.

\bibitem{neutel2002stability}
A.-M. Neutel, J.~A. Heesterbeek and P.~C. De~Ruiter,
\newblock \emph{Stability in real food webs: weak links in long loops},
\newblock Science \textbf{296}(5570), 1120 (2002),
\newblock \doi{10.1126/science.1068326}.

\bibitem{jacquet2016no}
C.~Jacquet, C.~Moritz, L.~Morissette, P.~Legagneux, F.~Massol, P.~Archambault
  and D.~Gravel,
\newblock \emph{No complexity--stability relationship in empirical ecosystems},
\newblock Nature communications \textbf{7}(1), 1 (2016),
\newblock \doi{10.1038/ncomms12573}.

\bibitem{moore2012energetic}
J.~C. Moore and P.~C. de~Ruiter,
\newblock \emph{Energetic food webs: an analysis of real and model ecosystems},
\newblock Oxford University Press,
\newblock \doi{10.1093/acprof:oso/9780198566182.001.0001} (2012).

\bibitem{baron2020dispersal}
J.~W. Baron and T.~Galla,
\newblock \emph{Dispersal-induced instability in complex ecosystems},
\newblock Nature communications \textbf{11}(1), 1 (2020),
\newblock \doi{10.5281/zenodo.4068257}.

\bibitem{fyodorov2016nonlinear}
Y.~V. Fyodorov and B.~A. Khoruzhenko,
\newblock \emph{Nonlinear analogue of the may- wigner instability transition},
\newblock Proceedings of the National Academy of Sciences \textbf{113}(25),
  6827 (2016),
\newblock \doi{10.3410/f.726406145.793528093}.

\bibitem{PhysRevE.103.022201}
S.~Belga~Fedeli, Y.~V. Fyodorov and J.~R. Ipsen,
\newblock \emph{Nonlinearity-generated resilience in large complex systems},
\newblock Phys. Rev. E \textbf{103}, 022201 (2021),
\newblock \doi{10.1103/PhysRevE.103.022201}.

\bibitem{bizeul2021positive}
P.~Bizeul and J.~Najim,
\newblock \emph{Positive solutions for large random linear systems},
\newblock Proceedings of the American Mathematical Society \textbf{149}(6),
  2333 (2021),
\newblock \doi{10.1090/proc/15383}.

\bibitem{10.21468/SciPostPhys.12.1.016}
S.~Franz, F.~Nicoletti, G.~Parisi and F.~Ricci-Tersenghi,
\newblock \emph{{Delocalization transition in low energy excitation modes of
  vector spin glasses}},
\newblock SciPost Phys. \textbf{12}, 016 (2022),
\newblock \doi{10.21468/SciPostPhys.12.1.016}.

\bibitem{fyodorov2021counting}
B.~Lacroix-A-Chez-Toine and Y.~Fyodorov,
\newblock \emph{Counting equilibria in a random non-gradient dynamics with
  heterogeneous relaxation rates},
\newblock arXiv preprint arXiv:2112.11250  (2021).

\bibitem{bun2017cleaning}
J.~Bun, J.-P. Bouchaud and M.~Potters,
\newblock \emph{Cleaning large correlation matrices: tools from random matrix
  theory},
\newblock Physics Reports \textbf{666}, 1 (2017),
\newblock \doi{10.1016/j.physrep.2016.10.005}.

\bibitem{wilkinson1988algebraic}
J.~H. Wilkinson,
\newblock \emph{The algebraic eigenvalue problem},
\newblock Oxford University Press, Inc.,
\newblock \doi{10.1017/S0013091500012104} (1988).

\bibitem{metz2021localization}
F.~L. Metz and I.~Neri,
\newblock \emph{Localization and universality of eigenvectors in directed
  random graphs},
\newblock Physical Review Letters \textbf{126}(4), 040604 (2021),
\newblock \doi{10.1103/PhysRevLett.126.040604}.

\bibitem{giorgio}
G.~Carugno, I.~Neri and P.~Vivo,
\newblock \emph{Instabilities of complex fluids with partially structured and
  partially random interactions},
\newblock Physical Biology \textbf{19}, 056001 (2022),
\newblock \doi{10.1088/1478-3975/ac55f9}.

\end{thebibliography}

\nolinenumbers

\end{document}